\documentclass[fleqn,usenatbib]{mnras}

\usepackage[T1]{fontenc}

\DeclareRobustCommand{\VAN}[3]{#2}
\let\VANthebibliography\thebibliography
\def\thebibliography{\DeclareRobustCommand{\VAN}[3]{##3}\VANthebibliography}

\usepackage{graphicx}	
\usepackage{amsmath}	
\usepackage{amssymb}	
\usepackage{pifont}     

\usepackage{xspace}	    
\usepackage{xcolor}	    

\newcommand{\orcid}[1]{\href{https://orcid.org/#1}{\includegraphics[width=10pt]{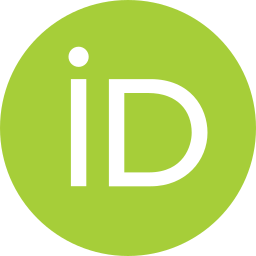}}}

\newcommand{\tick}{\hspace{1pt}\ding{51}}
\newcommand{\cross}{\hspace{1pt}\ding{55}}

\begingroup
\catcode`\_=\active
\gdef_#1{\ensuremath{\sb{\mathrm{#1}}}}
\endgroup
\mathcode`\_=\string"8000
\catcode`\_=12

\newcommand{\Msolar}{M$_{\odot}$\xspace} 
\newcommand{\HI}{H\textsc{i}\xspace}
\newcommand{\HII}{H\textsc{ii}\xspace}
\newcommand{\HeI}{He\textsc{i}\xspace}
\newcommand{\HeII}{He\textsc{ii}\xspace}
\newcommand{\HeIII}{He\textsc{iii}\xspace}
\newcommand{\rev}{\textcolor{black}}

\usepackage{newtxtext,newtxmath}

\title[Stellar Winds with Radiative Feedback]{The Energy and Dynamics of Trapped Radiative Feedback with Stellar Winds}

\author[Geen et al]{
Sam Geen$^{1,2}$\thanks{E-mail: s.t.geen@uva.nl} \orcid{0000-0002-3150-2543},
Rebekka Bieri$^{3}$ \orcid{0000-0002-4554-4488},
Alex de Koter$^{1,4}$  \orcid{https://orcid.org/0000-0002-1198-3167},
Taysun Kimm$^{5}$ \orcid{https://orcid.org/0000-0002-3950-3997},
Joakim Rosdahl$^{6}$ \orcid{0000-0002-7534-8314}
\\
$^{1}$ Anton Pannekoek Institute for Astronomy, Universiteit van Amsterdam, Science Park 904, 1098 XH Amsterdam, The Netherlands\\
$^{2}$ Leiden Observatory, Leiden University, PO Box 9513, 2300 RA Leiden, Netherlands \\
$^{3}$ Institute for Computational Science, University of Zurich, Zurich, Switzerland \\
$^{4}$ Institute of Astronomy, KU Leuven, Celestijnenlaan 200D, 3001 Leuven, Belgium \\
$^{5}$ Department of Astronomy, Yonsei University, 50 Yonsei-ro, Seodaemun-gu, Seoul 03722, Republic of Korea \\
$^{6}$ Univ Lyon, Univ Lyon 1, Ens de Lyon, CNRS, Centre de Recherche Astrophysique de Lyon UMR5574, F-69230 Saint-Genis-Laval, France \\
}

\date{Accepted XXX. Received YYY; in original form ZZZ}

\pubyear{2021}

\begin{document}
\label{firstpage}
\pagerange{\pageref{firstpage}--\pageref{lastpage}}
\maketitle

\begin{abstract}
In this paper we explore the significant, non-linear impact that stellar winds have on \HII regions.
We perform a parameter study using 3D radiative magnetohydrodynamic simulations of wind and UV radiation feedback from a 35 \Msolar star formed self-consistently in a turbulent, self-gravitating cloud, similar to the Orion Nebula (M42) and its main ionising source $\theta^1$ Ori C. 
Stellar winds suppress early radiative feedback by trapping ionising radiation in the shell around the wind bubble. Rapid breakouts of warm photoionised gas (``champagne flows'') still occur if the star forms close to the edge of the cloud. The impact of wind bubbles can be enhanced if we detect and remove numerical overcooling caused by shocks crossing grid cells. However, the majority of the energy in the wind bubble is still lost to turbulent mixing between the wind bubble and the gas around it. These results begin to converge if the spatial resolution at the wind bubble interface is increased by refining the grid on pressure gradients. 
Wind bubbles form a thin chimney close to the star, which then expands outwards as an extended plume once the wind bubble breaks out of the dense core the star formed in, allowing them to expand faster than a spherical wind bubble.  We also find wind bubbles mixing completely with the photoionised gas when the HII region breaks out of the cloud as a champagne flow, a process we term ``hot champagne''. 

\end{abstract}




\section{Introduction}

Massive stars strongly impact their environment via a set of processes called ``feedback'', terminating their own formation through a combination of protostellar feedback processes  \citep[e.g.][]{Kuiper2018,Bate2019}, clearing out their immediate formation environment \citep{Chevance2022a}, shaping the interstellar medium of galaxies \citep{McKee1977,Rathjen2021,Bieri2023} and even affecting the composition of the circumgalactic medium \citep{Tumlinson2017} and driving cosmic reionisation \citep[e.g][]{Rosdahl2018}. 

The problem of feedback in molecular clouds stems from both the complexity of the environment and the interaction between multiple feedback mechanisms. High energy radiation from stars plays a strong role in molecular clouds, both photoionising the gas around the stars, causing thermal expansion and dispersal of the cloud \citep{Kahn1954} and pushing the gas away via direct radiation pressure \citep{Mathews1967}.

Moreover, the strong ultraviolet radiation field of the star drives mass loss from the star as stellar winds, which can reach outflow velocities over $1000~$km/s for massive stars \citep{Puls2015}. This shock-heats the gas around the star, creating a hot, X-ray emitting wind bubble that expands into the surrounding medium \citep{Weaver1977}. 

Numerical simulations are a powerful tool for studying feedback because they self-consistently capture many of the complex features that emerge non-linearly from the interaction of different mechanisms in turbulent, magnetised, self-gravitating clouds. Simulations of photoionisation in a cloud context have been carried out by, e.g., \cite{Dale2005}, \cite{Gritschneder2009}, \cite{Peters2010}, among many others. Photoionisation-only feedback simulations with self-consistent star formation, e.g. \cite{Dale2012}, typically display rapid photoionisation of large portions of the cloud material followed by dispersal of the cloud due to thermal expansion. \cite{Ali2018} also reproduce this for a system similar to that in M42, and find some similarities between their simulations and M42 in, e.g. ionised gas velocities.

A key feature of photoionisation-driven cloud dispersal is the ``champagne'' flow \citep{TenorioTagle1979}, in which a density gradient in the cloud triggers a rapid breakout of ionised gas. This can occur due to a sharp step in density as the cloud edge is reached \citep{Whitworth1979} or due to a steep density gradient away from the star in the cloud itself \citep{Franco1990}. \cite{Comeron1997} also study this effect with winds included, which produces an embedded wind plume within the photoionised outflow.

Similarly, simulations of winds in molecular clouds have been carried out by, e.g. \cite{Rogers2013}, \cite{Rey-Raposo2017}, \cite{Geen2020} and \cite{Lancaster2021c}. \cite{Rosen2021} argue using a suite of simulations that winds alone can explain the sizes of feedback-driven shells around massive stars ($M > 10~$\Msolar), but not around lower mass stars. 

The combination of winds and radiation was explored by \cite{Dale2014}, who found that winds are relatively ineffective in contributing to stellar feedback in molecular clouds compared to photoionisation. This is also found in simulations of larger spiral arm regions by \cite{Ali2022}. Simulations by \cite{Grudic2022} and \cite{Guszejnov2022} establish that winds and radiation in concert effectively regulate star formation, though they also name protostellar jets as having a key role in both quenching star formation and in removing and disrupting matter accreting onto protostars. Supernovae generally occur too late to have a principal role in shaping star formation for molecular clouds with a global freefall time $\lesssim 7~$Myr.

\subsection{Feedback Efficiency and Radiative Cooling}

The efficiency of stellar feedback depends not only on the total quantity of energy deposited into the interstellar medium by stars, but also the ability for the interstellar medium to retain this energy as kinetic flows that impact larger scales. For example, \cite{Walch2012} argues that typically less than 0.1\% of the energy from ionising radiation is converted to kinetic energy in the gas via photoionisation. This is because while this process is important in thermalising the gas and driving outflows, most of the energy in photons is lost to radiative cooling of the gas at lower photon energies. The cooling rate becomes drastically higher in denser gas, particularly in very dense molecular clouds such as those in the central molecular zone \citep{Barnes2020} or in dense shells swept up by feedback \citep[e.g.][]{Rahner2017,Geen2022}. While ionising radiation can have an additional impact via direct radiation pressure, this is typically weak at \HII region radii more than a few pc \citep[e.g.][]{Oliver2021}.

Stellar winds rely on the overpressurisation of the hot bubble shocked by the stellar wind to drive expansion of the feedback-driven structures. The energy retained inside the wind bubble, which is mostly thermal energy, is roughly 45\% for an adiabatic wind bubble in a uniform medium \citep{Weaver1977} or 33\% in a singular isothermal sphere density profile ($\rho \propto r^{-2}$, see \citealp{Geen2022}). The rest is used to do thermodynamic work on the gas swept up by the wind bubble.

There are various channels via which the pressure in the wind bubble can be lost. Firstly, thermal evaporation of material from the dense wind-driven shell can cause mixing between hot, diffuse gas in the wind bubble and cold, dense gas, which enhances radiative cooling \citep{MacLow1988}. \cite{Kruijssen2019} argues that such cooling is inefficient on the timescales for feedback in molecular clouds. Thermal evaporation can be enhanced by thermal conduction via electrons \citep{Spitzer1962}, which \cite{Fierlinger2016} finds reduces the energy in a typical wind bubble by 10\% when running 1D simulations that include it. Finally, turbulent mixing between the hot, wind-shocked gas and denser gas outside (whether a cold, dense shell or a warm, moderately dense photoionised region) can be highly effective at enhancing cooling in the wind bubble. Early work by \cite{Kahn1980} suggested that Kelvin Helmholz instabilities in the wind bubble cannot grow fast enough to encourage strong cooling. However, more recent observational analysis \citep{Rosen2014} and analytic work \citep{Lancaster2021a} confirmed using simulations \citep{Lancaster2021b}, as well as work on shocks in more generalised environments \citep{Fielding2020,Tan2021} find that turbulent mixing, particularly in the context of structured or turbulent cloud environments, can effectively remove nearly all of the energy injected by stellar winds from the wind bubble, causing it to expand in a momentum-driven rather than pressure-driven mode. This is also found in simulations that include both stellar winds and photoionisation \citep{Dwarkadas2022}.

However, numerical simulations have significant limitations in reproducing all of these mechanisms. Spatial resolution, which is at a premium for 3D simulations, limits how effectively each of these processes can be captured given extant limits on computational power \citep{Pittard2021}. The inclusion of winds creates an additional need to resolve the short timesteps demanded by including hot and/or fast wind-driven flows, with characteristic speeds exceeding $1000~$km/s.

Additionally, certain numerical features of simulation codes can cause problems in accurately resolving the energetics of hot structures in the interstellar medium. One consequence specific to Eulerian hydrodynamic codes (i.e. codes that use static computational elements to simulate moving flows) is that shocks must cross cells. In doing so, the contact discontinuity between the hot, shocked gas and cold, unshocked gas must exist inside a cell or set of cells. \cite{Gentry2017} show that, in the case of supernova-driven superwind bubbles, the artificial mixing of hot and cold gas that occurs in these cells causes considerable numerical overcooling compared to simulations using Lagrangian hydrodynamics (i.e. codes that use computational elements that move with flows, hence reducing while not completely eliminating the effect of this artificial mixing). It is thus crucial to understand in detail how the energy from feedback transfers to gas on larger scales if we are to have an accurate picture for how stars shape the wider universe.

\rev{The Orion region provides an excellent observational case study for feedback in a nearby star-forming region. Star formation has been ongoing for up to 10 Myr in this region \citep{Alves2012,DaRio2012,Drass2016}, with evidence of ongoing massive stellar feedback in the Orion-Eridanus superbubble, powered the OB1 cluster's massive stars and supernovae \citep{Brown1994}. In this paper we focus on the case study of the young M42 nebula in this region, powered by the Trapezium cluster, and in particular by the O7V star $\theta^1$ Ori C, with secondary contributions from the O9.5V star $\theta^2$ Ori A \citep{ODell2017}. \cite{SimonDiaz2006} analyse this cluster and find an age of $2.5\pm0.5~$Myr for the cluster as a whole, although they urge caution given the uncertainties in measuring the age of young stars where initial rotation velocities are poorly understood. By contrast, \cite{Pabst2019} analyses the velocity structure of the M42 nebula itself using 3D spectroscopic observations of C$^{+}$, finding an age of 0.3 Myr based on an analysis encompassing the expansion rate and spatial size of the nebula. Interpreting the age of young star-forming regions precisely remains a challenge, and the input from theoretical models of both stars and nebulae remains important in piecing together the history of our Galactic neighbourhood.}

The goal of this paper is to assess how well numerical simulations capture the evolution and properties of HII regions powered by massive stars by focussing on the example of a single star in conditions similar to the solar neighbourhood \rev{and in particular M42 in Orion}, where we have a wealth of high resolution multiwavelength and spectroscopic observations. Rather than exploring a parameter space of stellar properties as we do in \cite{Geen2020}, we explore a set of numerical choices that affect the evolution of the wind bubble in the simulation, and whether we can recover a consistent set of behaviours for the wind bubble. These are grid refinement on pressure gradients (ensuring that interfaces in the \HII region are well resolved), masking the numerical cooling described above, the inclusion or otherwise of radiative feedback, and the random seed used to generate the cloud initial conditions.


\subsection{Paper Overview}

We run a series of simulations of wind bubbles around a self-consistently formed star, similar to the star $\theta^1$ Ori C that powers the Orion Nebula M42, in a cloud similar to Orion, which has been the focus of multiple recent multi-wavelength observational studies. We run these simulations with varying parameters, described above, in order to assess whether we can reproduce the broad observed features of the Orion nebula, and hence motivate further more detailed studies.

We begin by laying out the methods used in Section \ref{methods}. We then discuss the results in Section \ref{results}, focussing on the combination of wind and radiative feedback, the physics at the interfaces between the wind bubble, the photoionised \HII region and the neutral cloud, and the role of randomness in setting the initial conditions. Finally, in we present our conclusions in Section \ref{conclusions}.

\section{Methods}
\label{methods}

We simulate a set of isolated molecular clouds with an initial magnetic and turbulent velocity field, self-gravity and stellar feedback. The simulations are performed using the radiative magnetohydrodynamic (RMHD) Eulerian Adaptive Mesh Refinement (AMR) code \textsc{ramses} \citep{Teyssier2002,Fromang2006,Rosdahl2013}. Details of the full data reproduction package are available in Section \ref{data_availability}. 

The simulations are similar to the setup described in \cite{Geen2020}, in which we simulated a set of massive stars between 30 and 120$~$\Msolar using one set of initial conditions. The main differences in the physics implemented are the full refinement of cells across steep pressure gradients, and a module that identifies and masks out cooling in cells across the wind bubble contact discontinuity. The goal of this is to explore in depth whether these numerical choices affect the resulting \HII region properties, particularly when used with different combinations of feedback processes from the star. Full details are given below for completeness.

We list a summary of simulations used in this project in Table \ref{methods:simtable}. We divide our suite of simulations into different labelled sets, used to identify a specific parameter space exploration. The simulations in this study form part of the \textsc{pralines} suite \citep{kimm2022}, with the wider goal of understanding how radiative and mechanical feedback propagate to larger, (extra-)galactic scales. 

In the rest of this Section we discuss the numerical methods used to construct these simulations and the physical models employed.

\begin{table}
	\centering
	\caption{List of simulations included in this paper according to the set they are included in.  See Section \ref{methods} for more detail concerning the setup of the simulations.  UV means that UV photoionisation is included. Direct radiation pressure is also included in the UV runs except in the UV Only runs - see Section \ref{methods:stellar_feedback} for a discussion of this choice. WIND means that stellar winds are included. MASK means that the cooling mask at the contact discontinuity is included. REFINE means that refinement on pressure gradients is included (refinement on spatial coordinates, the Jeans criterion and around the stellar source are always included). SEED refers to the initial random seed used to initialise the turbulent velocity field, described in Section \ref{methods:initial_conditions}. As these seeds are not ordered sequentially in a meaningful way, we give each of them an arbitrary name to reference them throughout this work. Notes: $^{1}$These simulations are identical, and labelled differently in each set to clarify how they differ within the set. $^{2}$This seed was used in \protect\cite{Geen2020}, without the cooling mask or pressure gradient refinement turned on. $^{3}$This simulation is used in the ``hot champagne'' analysis.}
	\label{methods:simtable}
	\begin{tabular}{lccccccc} 
		\hline
		Simulation Name      & UV     & WIND   & MASK   & REFINE & SEED      \\
		\hline
		Fiducial Run$^{1}$   & \tick  & \tick  & \tick  & \tick  & \textsc{Bellecour} \\
		\hline
		Set \textsc{wind only} \\
		\hline
		Wind Only, Mask On   & \cross & \tick  & \tick  & \tick  & \textsc{Bellecour} \\
		Wind Only, Mask Off  & \cross & \tick  & \cross & \tick  & \textsc{Bellecour} \\
		\hline
		Set \textsc{feedback} \\
		\hline
		No Feedback          & \cross & \cross & \cross & \cross & \textsc{Bellecour} \\
		Wind Only            & \cross & \tick  & \tick & \tick   & \textsc{Bellecour} \\
		UV Only              & \tick  & \cross & \cross & \tick  & \textsc{Bellecour} \\
		Wind \& UV$^{1}$     & \tick  & \tick  & \tick & \tick   & \textsc{Bellecour} \\
		\hline
		Set \textsc{physics} \\
		\hline
		No Mask, No Refine   & \tick  & \tick  & \cross & \cross & \textsc{Bellecour} \\
		No Mask, Refine      & \tick  & \tick  & \cross & \tick  & \textsc{Bellecour} \\
		Mask, No Refine      & \tick  & \tick  & \tick  & \cross & \textsc{Bellecour} \\
		Mask, Refine$^{1}$   & \tick  & \tick  & \tick  & \tick  & \textsc{Bellecour} \\
		\hline
		Set \textsc{seeds} \\
		\hline
		Seed \textsc{Ampere}$^{2}$    & \tick  & \tick  & \tick  & \tick  & \textsc{Ampere}    \\
		Seed \textsc{Bellecour}$^{1}$ & \tick  & \tick  & \tick  & \tick  & \textsc{Bellecour} \\
		Seed \textsc{Carnot}$^{3}$    & \tick  & \tick  & \tick  & \tick  & \textsc{Carnot}    \\
		Seed \textsc{Duhamel}         & \tick  & \tick  & \tick  & \tick  & \textsc{Duhamel}   \\
		Seed \textsc{d'Enghien}       & \tick  & \tick  & \tick  & \tick  & \textsc{d'Enghien} \\
		\hline
	\end{tabular}
\end{table}

\subsection{Initial Conditions}
\label{methods:initial_conditions}

Our initial conditions use the same setup as the \textsc{diffuse} cloud in \cite{Geen2020}, since it matches closely the gas density distribution found in Orion \citep{Geen2017}. We impose an initially spherically symmetric density field
\begin{align}
    \rho(r) ={}& \rho_0 / (1 + (r / r_c)^2) & r < 3\,r_c \nonumber \\
    \rho(r) ={}& 0.01 \rho_0 & 3\,r_c < r < 6\,r_c \nonumber \\
    \rho(r) ={}& \rho_{ISM} & r > 6\,r_c
\end{align}
where $\rho_0=1.80\times10^{-21}~$g$~$cm$^{-3}$ (equivalent to a hydrogen number density  $n_0=1078~$cm$^{-3}$), $r_c = 2.533~$pc and $\rho_{ISM}=2.24\times10^{-24}~$g$~$cm$^{-3}$ (equivalent to a hydrogen number density  $n_{ISM}=1.34~$cm$^{-3}$). The initial temperature is set to $10~$K inside $6\,r_c$ and $8000~$K outside. The simulation volume is a cubic box of length $24\,r_c = 60.8~$pc. The total mass inside $6\,r_c$ from the centre of the box is $10^4~$\Msolar.

We set up an initial magnetic field, oriented in the $x$ direction. The magnetic field strength along each line of sight in this direction is given by
\begin{equation}
    B_x = B_{max,ini} (\Sigma_x / \Sigma_{max,ini})
\end{equation}
where $B_{max,ini} = 3.76~\mu$G is the peak initial magnetic field strength, $\Sigma_x$ is the gas column density in the $x$ direction and $\Sigma_{max,ini}$ is the peak column density. This is a nominal initial magnetic field strength corresponding to an initial Alfv\'en crossing time to freefall time ratio of 0.2. The magnetic field strength evolves over time as density structures form in the cloud.

We set an initial turbulent velocity field in the cloud following the prescription given in \cite{Geen2020} at $t=0$, with no further turbulent forcing applied. The turbulent velocity field's root-mean sphere velocity, $v_{RMS}$ is nominally set to give a crossing time of half the free-fall time. \rev{The turbulent velocity field is created by convolving a white noise field with a Kolmogorov power law. The white noise field is initialised with a given random seed. A list of the seeds used to generate specific white noise fields, and hence repeatable sets of initial conditions, is given in Table \ref{methods:simtable}. All of the initial conditions generated by these seeds are created identically except for the value of the seed.}

\subsection{Initial Evolution and Refinement}

We evolve the simulation using the \textsc{hlld} solver for \textsc{mhd} \citep{Miyoshi2005} as described in \cite{Fromang2006}.

The average free-fall time of the gas inside $3\,r_c$ is $4.22~$Myr. We allow the cloud to evolve without self-gravity for half of this time (2.1 Myr) in order to allow the density and turbulent velocity field to mix \citep[see][]{Klessen2000}. After this time we turn on self-gravity. 

We use a cubic octree mesh that refines adaptively under defined conditions. Each refinement level involves a cell subdividing $2^3$ times. We set a coarse cubic grid of $2^7 = 128$ cells per side across the simulation volume. We further fully refine the grid up to level 9 inside a sphere of diameter $12\,r_c$. 

Our maximum refinement level is 11, giving a minimum cell length of $\Delta x = 0.03~$pc. This can be reached under three conditions. Firstly, we refine any gas that is denser than ten times the Jeans density, $\rho_J \equiv (c_s/\Delta x)^2 / G$, where $c_s$ is the sound speed in the cell. Secondly, we refine any cell more massive than $0.25~$\Msolar. Thirdly, we refine any cell whose neighbour has a 99\% difference in pressure (i.e. $2 |P_{left}-P_{right}|/(P_{left}+P_{right}) > 0.99$. Simulations without this third refinement criterion are labelled ``No Refine'', however all simulations including those labelled as such include the first two types of refinement.

\subsection{Radiative Transfer}
\label{methods:radiativetransfer}

We use fully coupled multigroup radiative transfer with the M1 method. Full details of the method as described here are given in \cite{Rosdahl2013}. Photons interact with the gas via photoionisation, dust absorption and direct radiation pressure. The handling of sources of stellar radiation are discussed in Section \ref{methods:stellar_feedback}.

We track photons in groups representing far ultraviolet (FUV, between 11.2 and 13.6 eV) and 3 extreme ultraviolet (EUV) groups capable of ionising \HI to \HII (between 13.6 eV and 24.59 eV), \HeI to \HeII (24.59 eV to 54.4 eV) and \HeII to \HeIII (above 54.4 eV). All photons in each group are assumed to have the same energy and cross-sections.

We store the photon density and flux for each group in each cell. We use a reduced speed of light of 0.01 c, designed to capture high ionisation front speeds. The radiation field can evolve multiple times in one (magneto)hydrodynamic timestep (termed ``subcycling'') if the radiation timestep is shorter than the MHD timestep, each governed by the typical crossing time of radiation and gas flows in a cell. 

Each gas cell tracks the ionisation state of hydrogen and helium. Gas becomes ionised through photoionisation and collisional ionisation, and becomes neutral through recombination. 

We implement radiation pressure on gas and dust as described in \cite{Rosdahl2015}. The local gas opacity to absorption ($\kappa_{abs}$) for all EUV groups, and scattering ($\kappa_{sc}$) for all groups, is given by $10^3 Z / Z_{ref}~$cm$^2$/g, where Z is the metallicity of the gas ($=0.014$ in all runs, matching the Solar metallicity used in \citealp{Ekstrom2012}) and $Z_{ref}=0.02$. We do not track the creation and destruction of dust as a separate fluid, as in, e.g., \cite{Lebreuilly2019}. Instead, we approximate the effect of dust by including an additional absorption and scattering opacity, which are given by
\begin{equation}
    \left\{\kappa_{abs},\kappa_{sc} \right\}_{dust} = \left\{\kappa_{abs},\kappa_{sc} \right\}_{gas} \left(\frac{T_R}{10~\mathrm{K}}\right)^2 \, \mathrm{exp}\left(\frac{-T_R}{10^3~\mathrm{K}}\right) (1 - x_{HII}) \rho,
\end{equation}
where ``abs'' refers to absorption and ``sc'' refers to scattering. The term $10^3~$K is designed to mimic dust destruction by sputtering above this temperature. $T_R = (E_R / a)^{1/4}$ is the local radiation temperature, where $E_R$ is the energy of the photons in each group in the cell and $a \equiv 7.566\times10^{-15}~$erg$~$cm$^{-3}~$K$^{-4}$ is the radiation constant, $x_{HII}$ is the local hydrogen ionisation fraction and $\rho$ is the local gas density. A full dust model coupling to radiation pressure is left to future work, though based on our previous models in \cite{Geen2020} and \cite{Geen2022}, we do not expect this to change our results significantly for the systems studied here.

\subsection{Radiative Cooling}

In addition to radiative transfer of high energy photons, a certain fraction of the gas thermal energy is considered lost to low-energy radiation in each timestep. Hydrogen and helium follow the cooling and heating functions described in \cite{Rosdahl2013}. Metal cooling is accounted for with separate functions following the neutral and ionised hydrogen fractions in each cell. Metal cooling for the neutral fraction of a cell is given by the model of \cite{Audit2005} which tracks cooling of various elements in the ISM. Metal cooling in the ionised fraction uses a piecewise fit to the cooling curve given in \cite{Ferland2003}, with a fit to \cite{Sutherland1993} above $10^4~$K. For all of our simulations, the metal fraction is $Z=0.014$, i.e. Solar metallicity. We ignore metal enrichment by stellar mass loss as we expect this to be small compared to the initial cloud metallicity.

Shocks moving on Eulerian grid codes are well captured when adiabatic, but when cooling is included, artificial numerical mixing occurs as the shock crosses a cell and a contact discontinuity exists between the hot, diffuse gas post-shock and the cold, dense gas pre-shock. This creates an artificially warm, dense cell that leads to spurious overcooling. This is discussed in some detail in \cite{Fierlinger2016} and \cite{Gentry2017}, the latter for the case of supernova remnants rather than main sequence stellar winds.

In order to mitigate this effect, we invoke a very simple fix that masks cells in which neighbouring cells (identified in a 3x3x3 cube centred on a given cell) contain both cells above $10^6~$K and cells below $10^6~$K, and turns off cooling in this mask. We use $T=10^6~$K as a reliable heuristic for whether gas is in the wind-shocked bubble or outside it. We discuss the implications of our simplified choice later in the paper. Our simulations are considered to have this fix by default, or are labelled ``Mask''. Simulations without this contact discontinuity mask are labelled ``No Mask''.

\subsection{Sinks and Star Formation}

In our simulations, we follow star formation using sink particles that accrete from the surrounding gas. We first identify if cells reach 10\% of the Jeans density at the highest level of refinement. We then identify peaks amongst these cells using the watershed method described in \cite{Bleuler2014}. If the peak is denser than the Jeans density at the highest level of refinement, we form a sink particle. Every timestep, this sink accretes 90\% of the mass in the clump above the Jeans density. For a complete description, see \cite{Bleuler2014a}. 

We use these sink particles to track material that collapses to scales smaller than those we can track with the AMR grid, and that become the sites of star formation. We consider a small cluster to have formed once 120 \Msolar has accreted in total onto sink particles. At this point, we create a stellar object and assign it to the most massive sink at the time of formation. This object is not in itself a N-body particle and rather sits on the position of this sink at all times. 

We track the age of the object from the moment of its creation and consider it to be a star of 35 \Msolar, similar to $\theta^1~$Ori$~$C, the source of the Orion Nebula, M42 \citep{Kraus2009,Balega2014}. This mass is used by the stellar evolution track to extract accurate wind and radiation emission rates for a given star in each timestep, and does not have any influence on the gravity in the simulation aside from the mass of the sink it is a part of. We do not allow any other stellar objects to form that produce feedback, in order to isolate the feedback from one star, although we allow sinks to continue accreting material, modelling the ongoing formation of lower-mass stars in the cloud. This is representative of the situation in M42. Future work will include feedback from multiple self-consistently forming stars, as in \cite{Geen2018}.

\subsection{Stellar Feedback}
\label{methods:stellar_feedback}

The age and mass of the stellar object is used to track the radiation and winds emitted by the star using a stellar feedback module. A full description of the feedback module with figures showing the tracks is given in \cite{Geen2020}. We summarise the module here.

We do not consider the pre-main sequence phase of the 35 \Msolar star since we do not track the scales relevant to protostar formation. The star is considered to immediately arrive at the zero age main sequence once the stellar object is created.

Our star emits energy as radiation in the four modelled groups (1 FUV, 3 EUV) and experiences mass, momentum and energy loss from stellar winds. Our simulations end before the star should go supernova (at an age of 6.8 Myr in our model). 

We base our stellar evolution tracks on the Geneva Model \citep{Ekstrom2012,Georgy2012}. We use the rotating tracks, i.e. stars rotating at 40\% of their critical velocity. Stellar spectra are extracted from \textsc{starburst99} \citep{Leitherer2014} with the Geneva models as inputs.

Our wind model is similar to that in \cite{Gatto2017}. Mass loss rates are taken from the Geneva model, with a corrected wind terminal velocity calculated from the escape velocity as in \cite{Vink2011}. A full description of the model used is given in \cite{Geen2020}.

Radiation is deposited onto the grid in a single cell where the host sink of the star is located. This radiation then propagates outwards following the M1 method. 

Winds are deposited onto the grid as mass, momentum and energy in a thick shell between 3.5 and 5 cells in radius at the maximum level of refinement around the host sink of the star. We do this to ensure that the initial state of the feedback region captures swept-up material by the unresolved compact \HII region around the star, which in turn affects how radiation from the star is allowed to propagate into the surrounding medium at $t \sim 0$. We inject all of the energy from the wind as kinetic energy. 

Initially, the background density of the cells that the wind is injected into is high. This means that the flow thermalises, since the momentum added by the wind bubble is negligible. As the wind sweeps out a low-density bubble, the mass injected by the wind accounts for the majority of the mass inside the injection radius, and so the flow becomes kinetic. We thus begin to capture the free-streaming radius described in \cite{Weaver1977}.

We perform runs with both winds and radiation, winds only and radiation only (see Table \ref{methods:simtable}). Runs with radiation only  (``UV Only'') omit direct radiation pressure on gas and dust. This is because of a physical effect where radiation pressure creates a central empty cavity around the star \citep[see, e.g.][]{Draine2011}, leading to an artificially short timestep in this region as the simulation attempts to balance the pressure in the region. This causes the simulation to slow down considerably. This effect does not occur when stellar winds of any kind are included, as the ambient volume around the star becomes filled with free-streaming wind material, and so in runs with winds, radiation pressure is also included. In this way we consider the effect to be unphysical, but include the UV Only run nonetheless for the purposes of comparison with the runs including winds. In any case, we note that radiation pressure has a negligible effect on our results, in agreement with our previous simulation work \citep{Geen2020} and analysis of observed \HII regions of a similar size \citep{Oliver2021}.

Additionally, for the UV Only run, we disable refinement on pressure gradients, since the simulation attempts to refine over a large volume once the ``champagne flow'' phase begins. This causes the simulation to request excessive memory and CPU resources.


\section{Results}
\label{results}

In this Section we describe the results of combining wind and radiative feedback when using the numerical recipes implemented in this work. We first present the results of combining winds and radiative feedback using a fiducial set of these recipes, finding that winds are capable of trapping radiative feedback as described in \cite{Geen2022}. We then study the effect of correcting for numerical overcooling at the wind bubble contact discontinuity, studying in particular the role of turbulent mixing when this correction is applied. We then study how the initial random seed used to generate the initial conditions plays a strong role in the evolution of the wind bubble.

\subsection{Combining Winds and Radiative Feedback}
\label{results:combiningwindsandradiativefeedback}

We now compare the influence of stellar winds and UV radiation feedback in shaping \HII regions. To do this, we analyse the \textsc{feedback} set of simulations, which explores the effect of individual feedback processes included in the simulations. Specifically, we review the combined impact of UV radiation and winds on the cloud.

\begin{figure*}
    \centering
    \includegraphics[width=\hsize]{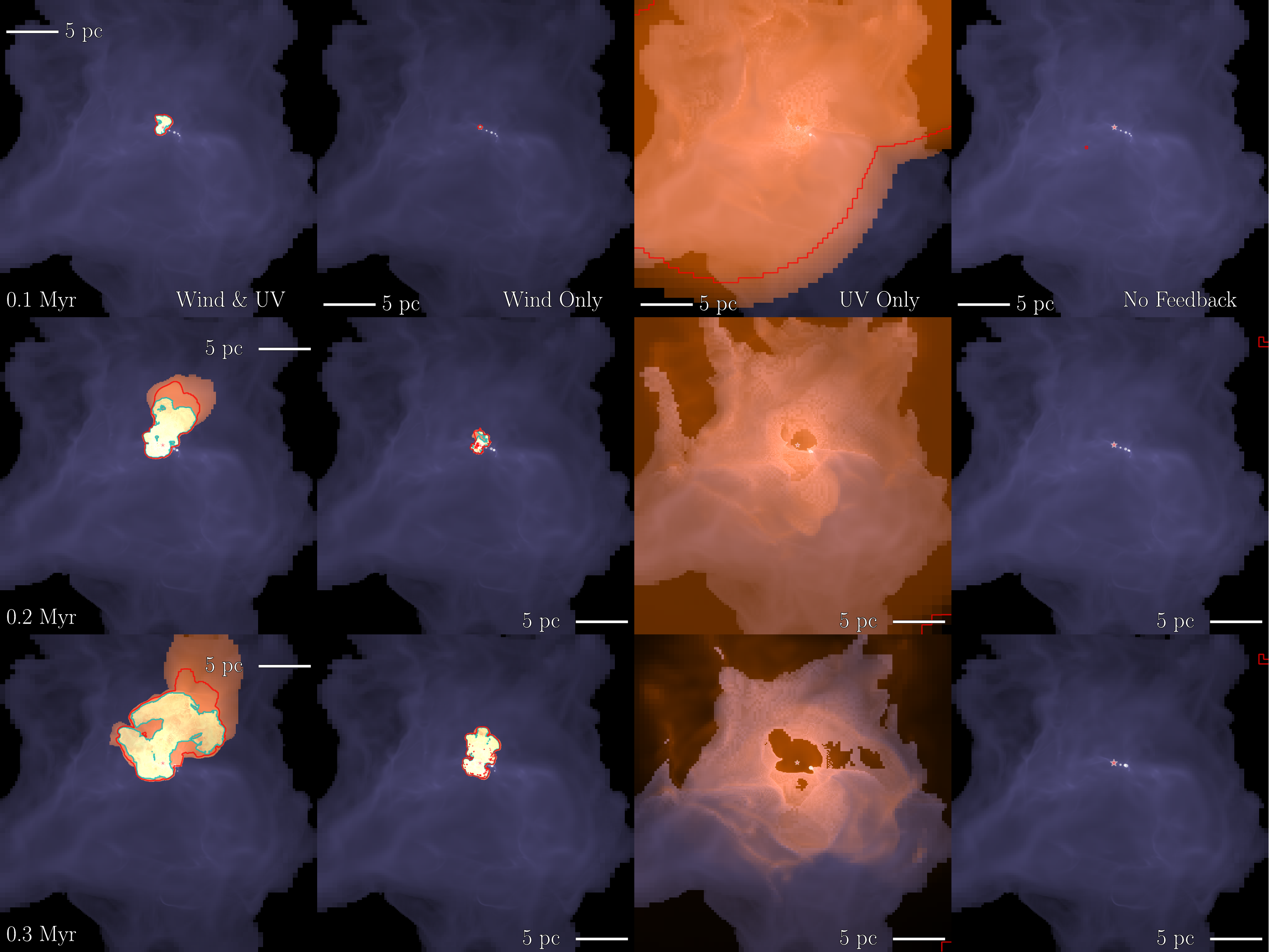}
    \caption{Sequence of normalised emission maps showing the evolution of the nebula around the star in the \textsc{feedback} set, where we vary whether winds or UV photoionisation are included. Each image shows a projection of emission from collisional excitation from cold gas ($\rho^2$ for cells where $T < 1000~$K), hot gas ($\rho^2$ for cells where $T > 106~$K), and recombination emission ($\rho^2 x_{HII}$, ignoring cells where $x_{HII} < 10^{-4}$). A cyan contour encloses pixels with gas above $10^6~$K in them. A red contour encloses pixels containing cells with a hydrogen ionisation fraction $x_{HII} > 0.1$, i.e. outlining where the extent of the \HII region lies. \rev{Some gas with a lower ionisation fraction is visible outside this region, as the ionisation front is not a sharp discontinuity in runs that exhibit champagne flows}. The columns from left to right show runs with winds and UV photoionisation, winds only, UV photoionisation only and no feedback.
	The rows from top to bottom show the outputs in each simulation at a stellar age of 0.1, 0.2 and 0.3 Myr. The apparent drop in recombination line emission at 0.3 Myr in the third column is due to normalisation effects and does not reflect a drop in photoionisation.}
    \label{fig:imagefb}
\end{figure*}

Figure \ref{fig:imagefb} shows a sequence of images of each simulation highlighting the presence of cold, neutral gas (purple), warm, photoionised gas (orange) and hot, wind-shocked gas (light yellow), with contours outlining the wind bubble (cyan) and photoionised gas (red) separating the phases of the \HII region around the star (red star icon), and with time running from top to bottom with snapshots at 0.1, 0.2 and $0.3~$Myr. 

\begin{figure}
    \centering
    \includegraphics[width=\columnwidth]{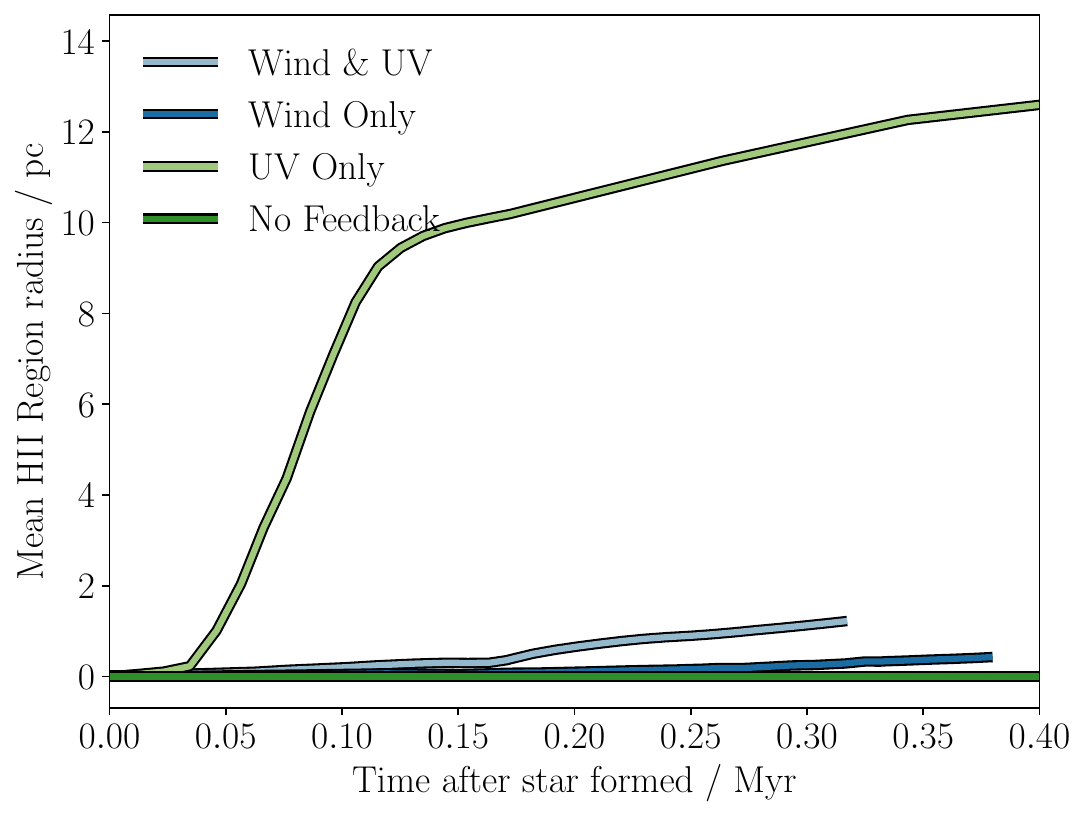}
    \caption{Mean spherical-equivalent \HII region radius $R_{HII}$ (including wind and photoionisation-heated gas, where the region volume $V_{HII} = \frac{4}{3} \pi R_{HII}^3$) as a function of time in each simulation in the \textsc{feedback} set. Winds act to constrain runaway photoionisation of the whole cloud by trapping the radiation in the shell around the wind bubble.}
    \label{fig:radiusfb}
\end{figure}

\begin{figure}
    \centering
    \includegraphics[width=\columnwidth]{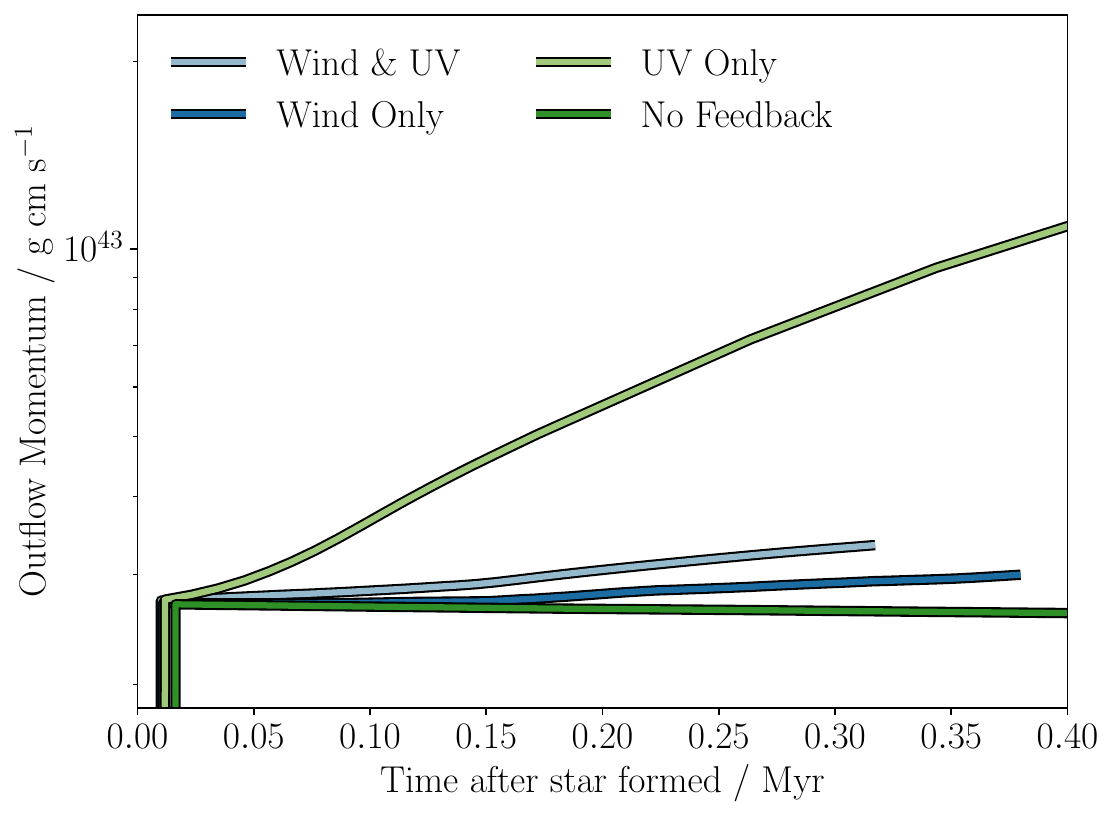}
    \caption{Momentum in flows directed away from the position of the star as a function of time in each simulation in the \textsc{feedback} set. As with the radial evolution in Figure \ref{fig:radiusfb}, including winds drastically reduces the impact of radiation feedback and hence the total effect of feedback on the cloud.}
    \label{fig:momentumfb}
\end{figure}

The notable result of this comparison is that combining winds and radiation from the star gives it a much weaker impact than simply including ionising radiation. This is also seen in the mean spherical-equivalent radial expansion of the region (Figure \ref{fig:radiusfb}, where the radius is calculated as $(3V_{II}/4\pi)^{1/3}$, where $V_{II}$ is the \HII region volume including both photoionised and hot, wind-shocked gas above $10^5~$K) and momentum in outflows as a result of feedback (Figure \ref{fig:momentumfb}, including only components of gas flows moving radially away from the star). In the UV Only run, there is a very rapid expansion of the \HII region at a stellar age of 40 kyr, corresponding to a ``champagne'' flow \citep{TenorioTagle1979,Whitworth1979}, in which the dense gas around the star cannot absorb all of the ionising radiation any more, causing the ionisation front to \rev{move outwards} supersonically without significant gas flows. This can be seen in the UV Only run in Figure \ref{fig:radiusfb}, where the mean radius of the \HII region expands rapidly after $\sim0.04~$Myr, before slowing at $0.1~$Myr where parts of the region reach the edge of the box. By comparison, the momentum of the gas in the UV Only run in Figure \ref{fig:momentumfb} \rev{advances} more slowly with no real sign of a slowing at $0.1~$Myr as the photoionised gas begins to respond hydrodynamically \citep[see][]{Franco1990}. 

The trapping of ionising radiation by a wind-blown shell in a power law density field is described analytically in \cite{Geen2022}, where the presence of a hot wind bubble acts to increase the pressure inside the \HII region, creating a denser shell around the region and hence trapping a large quantity of radiation. It is worth noting that this is created by swept-up material rather than the mass directly from the star, which is typically much smaller than the mass of the circumstellar medium in a dense cloud. Even when the champagne flow phase occurs in simulations where winds are included (around 0.17 Myr in the Wind \& UV simulation, seen in the increase in expansion rate in Figure$~$\ref{fig:radiusfb}), the ionisation front grows more slowly when winds are included. This suggests that a significantly reduced amount of radiation is able to escape even once the champagne phase begins. 

\begin{figure*}
    \centering
    \includegraphics[width=2\columnwidth]{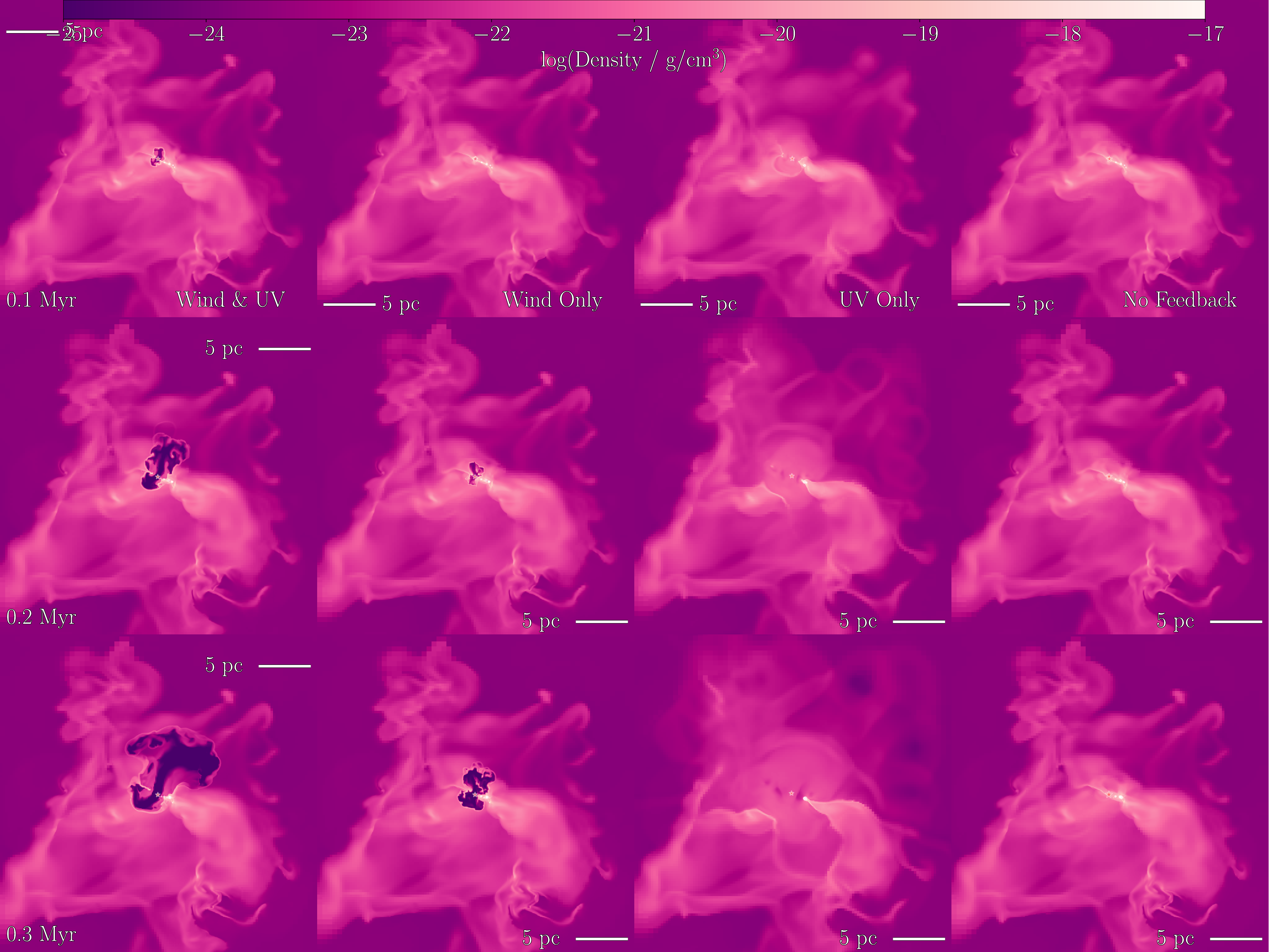}
    \caption{Sequence of slices through the position of the star showing gas density in the \textsc{feedback} set, where we vary whether winds or UV photoionisation are included. The columns show the same simulations as Figure \protect\ref{fig:imagefb}. Runs with winds contain dense shells around the wind bubble, which absorbs UV radiation from the star.}
    \label{fig:slicerhofb}
\end{figure*}

The presence of a dense shell can be seen by examining Figure$~$\ref{fig:slicerhofb}, where we plot slices through the position of the star showing gas density. In the UV Only run, the gas is rapidly photoionised as described in \cite{Franco1990}, and as such there is no shell around parts of the \HII region that form a Champagne flow. Rather, the photoionised cloud expands thermally. By contrast, although the champagne phase has begun in the Wind \& UV run in this figure, there is still a noticeable dense shell around the diffuse wind bubble that acts to absorb significant quantities of ionising radiation. 

There is some very limited leakage through parts of the shell in the Wind \& UV simulation. This can be seen when comparing the position of the red contour and the edge of the orange region in Figure \ref{fig:imagefb}, which use cut-off thresholds of $x_{HII} < 10^{-1}$ and $x_{HII} < 10^{-4}$ respectively. Despite this, nearly all of the ionising radiation is trapped, compared to the rapid champagne flow seen in the UV Only runs.

%
%
%

\subsection{The Physics at the \HII Region Interface}

A key component in \HII regions is the interface between different phases, in particular the contact discontinuity between the wind bubble and the gas around it, which is comprised of either a dense, partially- or a fully-photoionised shell of material.

To this end, we study two recipes designed to improve our simulations' ability to capture the physics of this region compared to \cite{Geen2020}, namely our masking out of cooling around the contact discontinuity between the wind bubble and the gas outside it (simulations labelled ``Mask'' when included, otherwise ``No Mask''), and the use of grid refinement to better resolve strong pressure gradients in our simulations (labelled ``Refine'' when pressure gradient refinement is included, or ``No Refine'' otherwise).


\begin{figure*}
	\centering
	\includegraphics[width=\columnwidth]{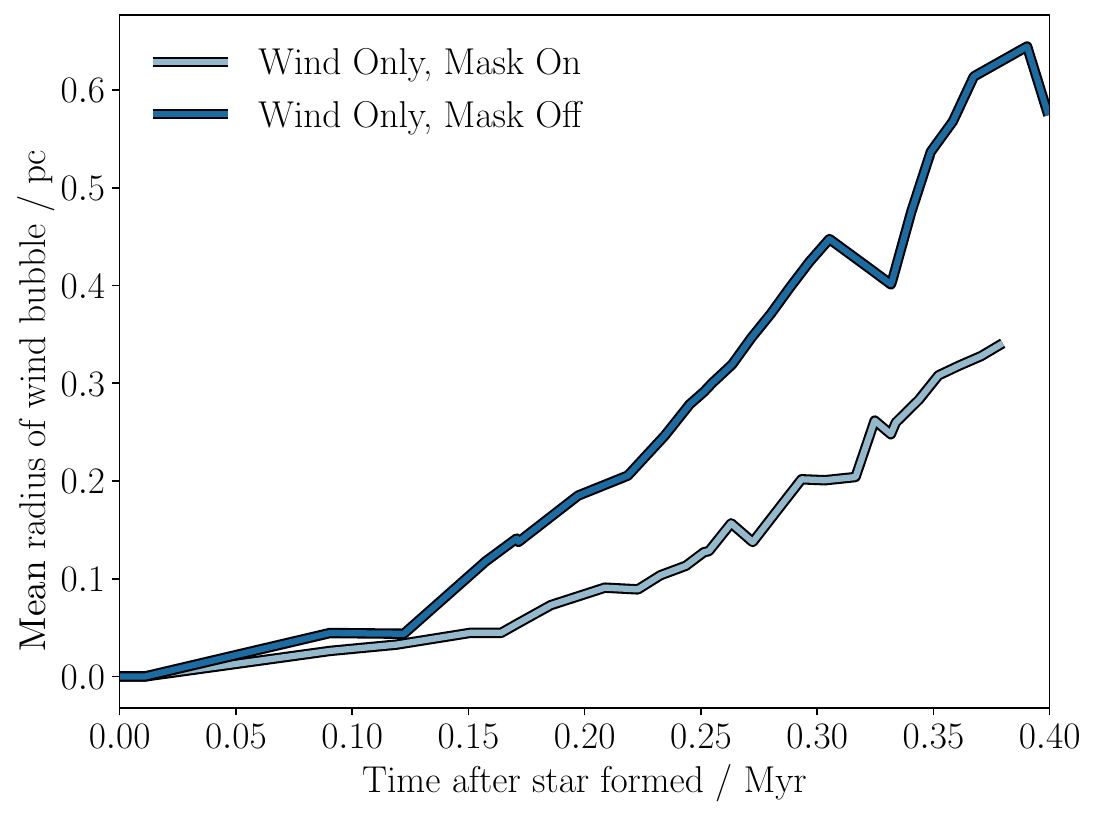}
	\includegraphics[width=\columnwidth]{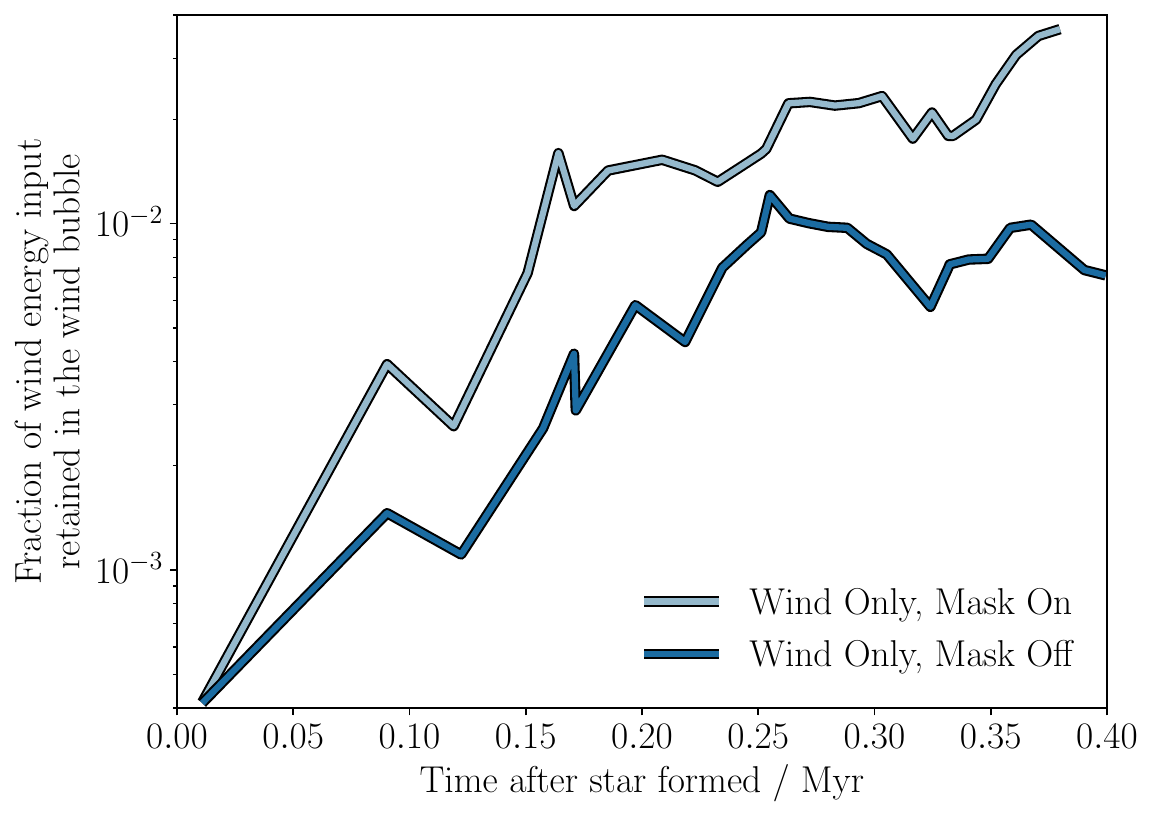}
	\caption{Comparison of runs in the \textsc{wind only} set where the cooling mask around the wind bubble contact discontinuity is either turned on or off. The left panel shows the mean spherical-equivalent radius of the wind bubble $R_w$ (where the wind bubble volume $V_{w} = 4/3 \pi R_w^3$, and $V_{w}$ is the volume of gas in cells above $10^5~$K). The right panel shows the fraction of energy from stellar winds retained in the wind bubble (kinetic + thermal). Despite retaining less energy, the wind bubble in the run without the cooling mask has a larger volume. We provide an explanation for this difference in Section \ref{results:windonlycomparison}.}
	\label{fig:windonlytest}
\end{figure*}

\begin{figure*}
	\centering
	\includegraphics[width=\hsize]{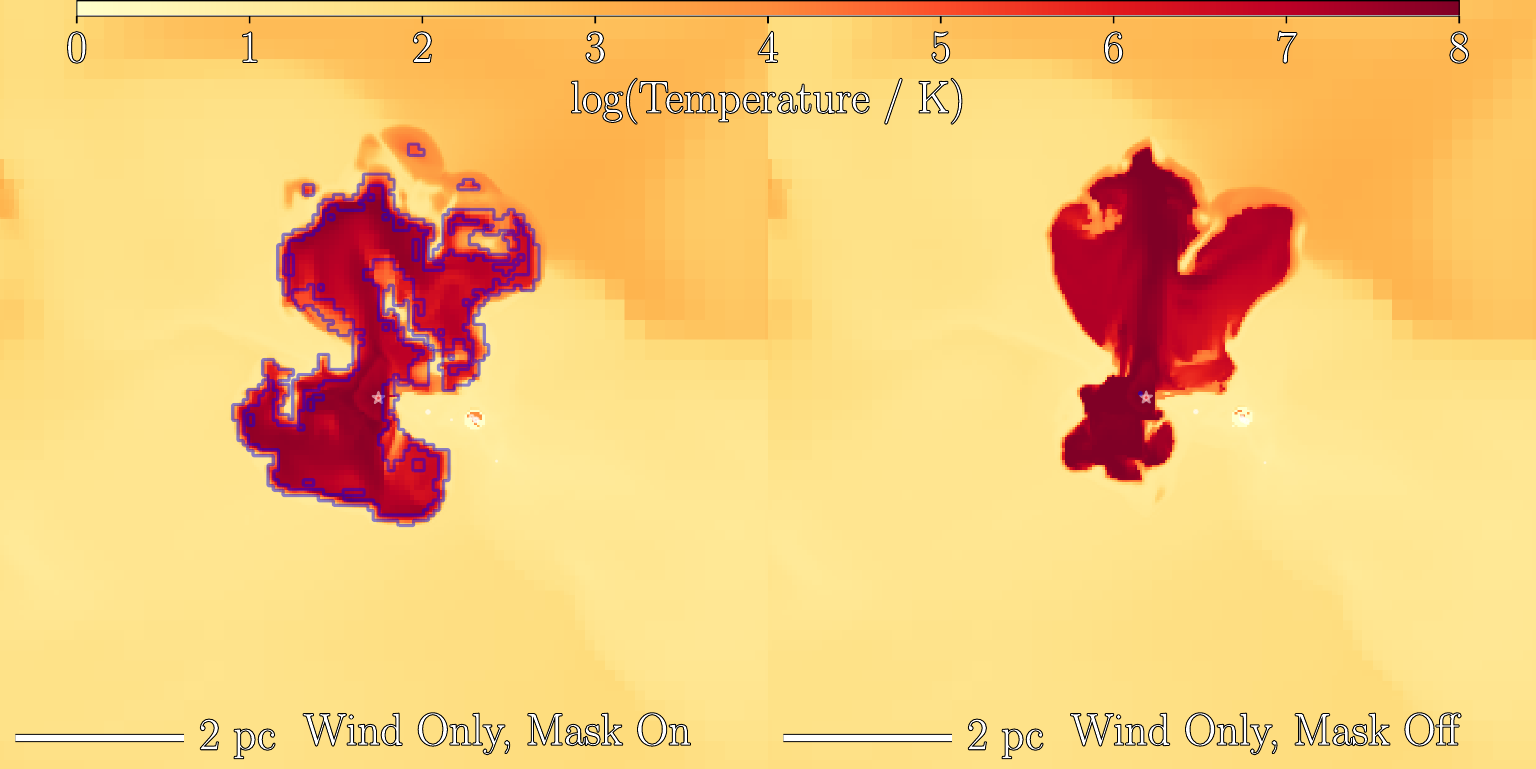}
	\caption{Slices in temperature through the position of the star in the \textsc{wind only} set at a stellar age of 0.3 Myr, comparing the effects of the cooling mask around the wind bubble contact discontinuity with only stellar winds included (no UV photoionisation). The left image with the cooling mask applied shows more internal substructure than the run shown on the right without the cooling mask. A translucent blue contour outlines the masked cells in the left image. The wind bubble is considered to be all gas cells above $10^5~$K, which does not include any substructure inside the wind bubble at a lower temperature.}
	\label{fig:windonlyslices}
\end{figure*}

\begin{figure*}
\centering
\includegraphics[width=\columnwidth]{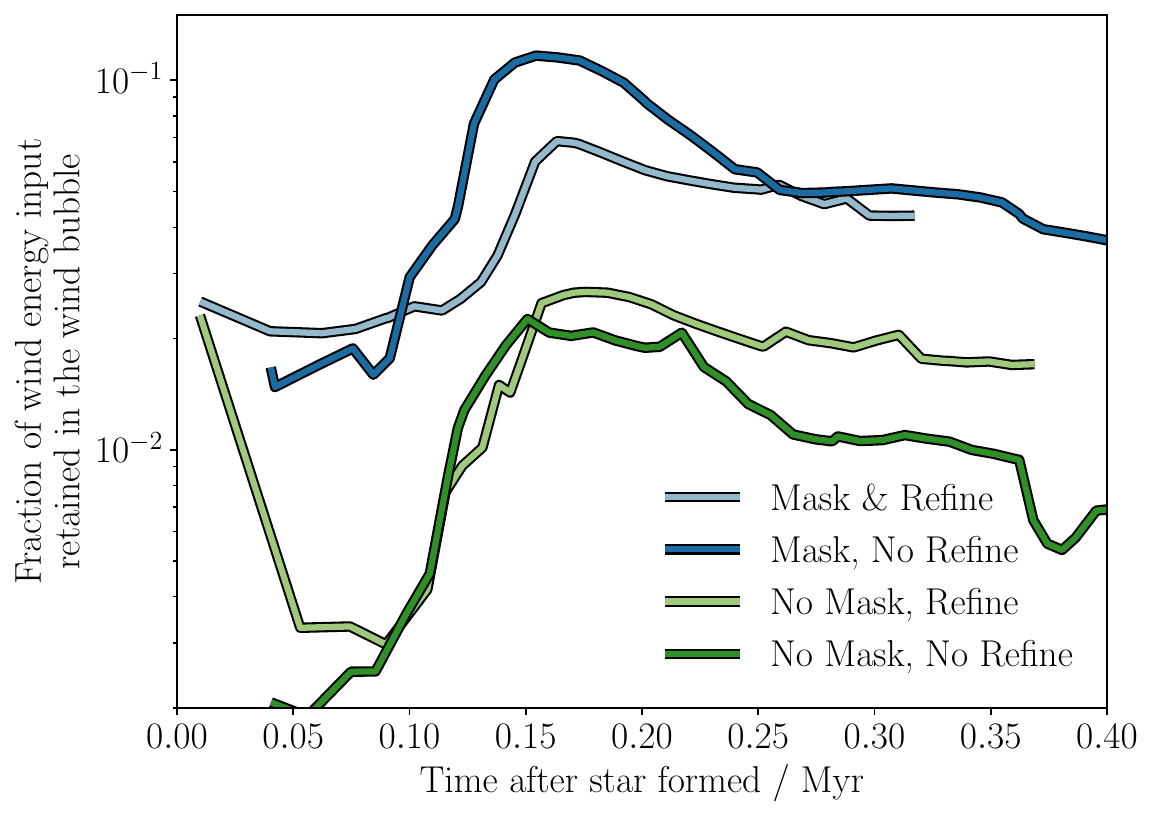}
\includegraphics[width=\columnwidth]{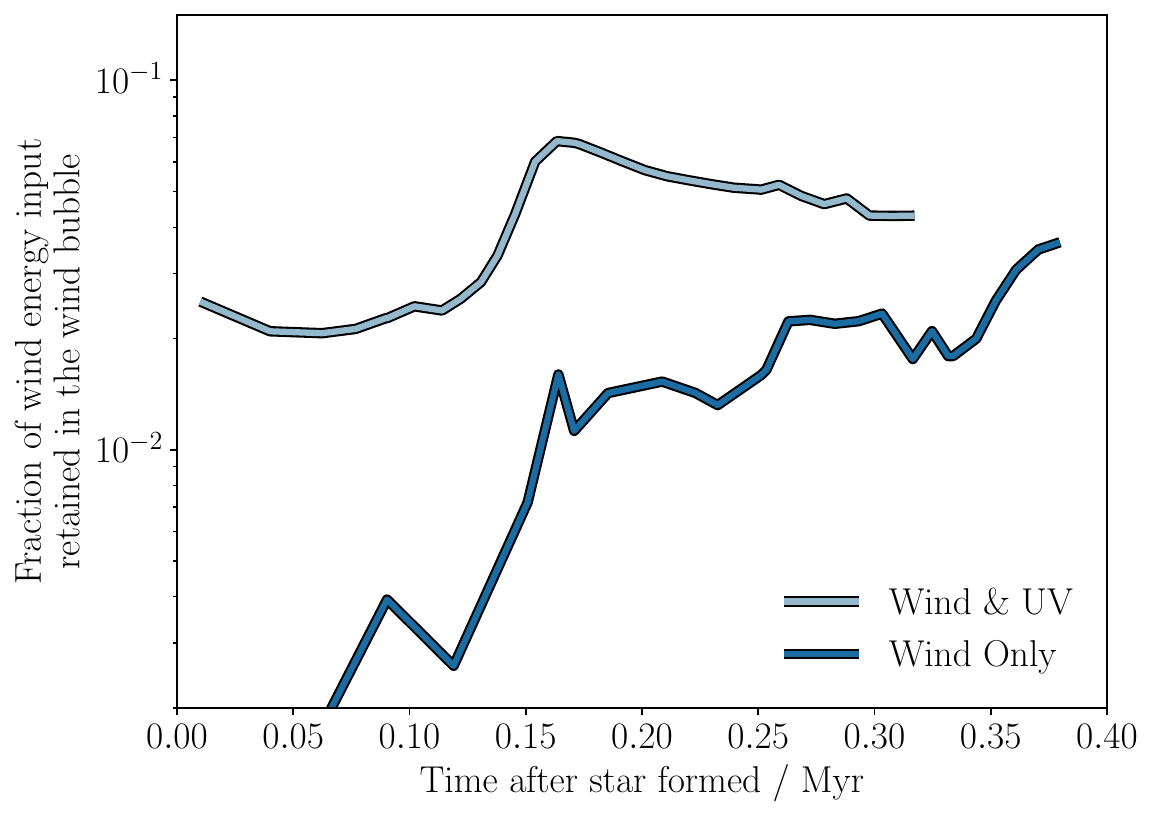}
\caption{The energy in the wind bubble as a fraction of the total energy injected as stellar winds, as a function of time in each simulation in the \textsc{physics} (left) and \textsc{feedback} (right) sets. Including UV photoionisation reduces wind cooling, as does including a cooling mask around the wind bubble contact discontinuity, though including refinement on pressure gradients causes some modest convergence between the results with and without the cooling mask.}
\label{fig:energyretainedphysicsfb}
\end{figure*}

\begin{figure*}
\centering
\includegraphics[width=\hsize]{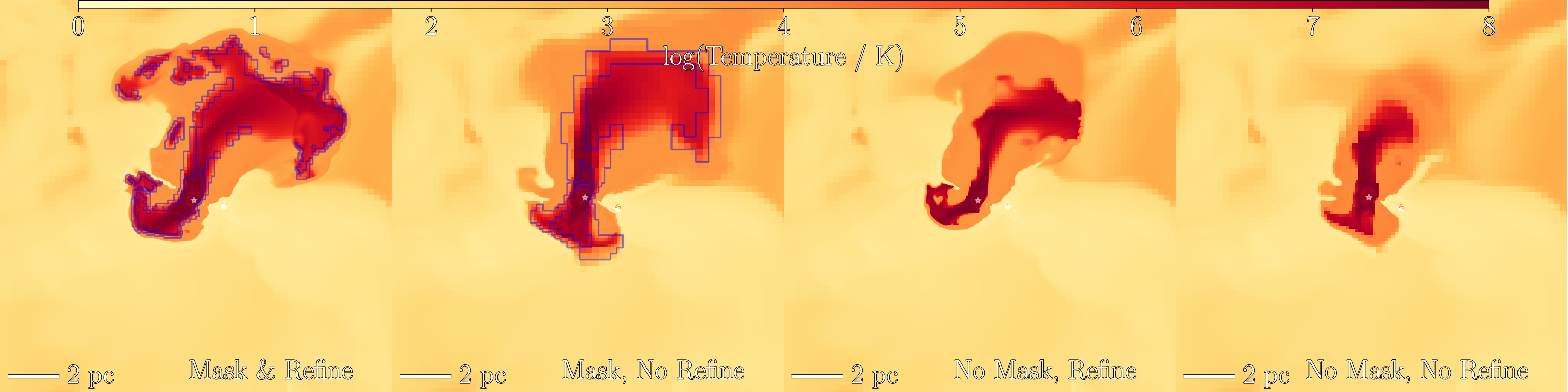}
\caption{Slices through the position of the stellar source showing gas temperature in the \textsc{physics} set at a stellar age of $0.3~$Myr. Gas above $10^5~$K is wind-shocked gas, gas at $\sim10^4~$K is photoionised and colder gas is neutral cloud material. A translucent blue contour outlines the masked cells in the runs that include a cooling mask.}
\label{fig:imagephysics}
\end{figure*}

\begin{figure*}
\centering
\includegraphics[width=\columnwidth]{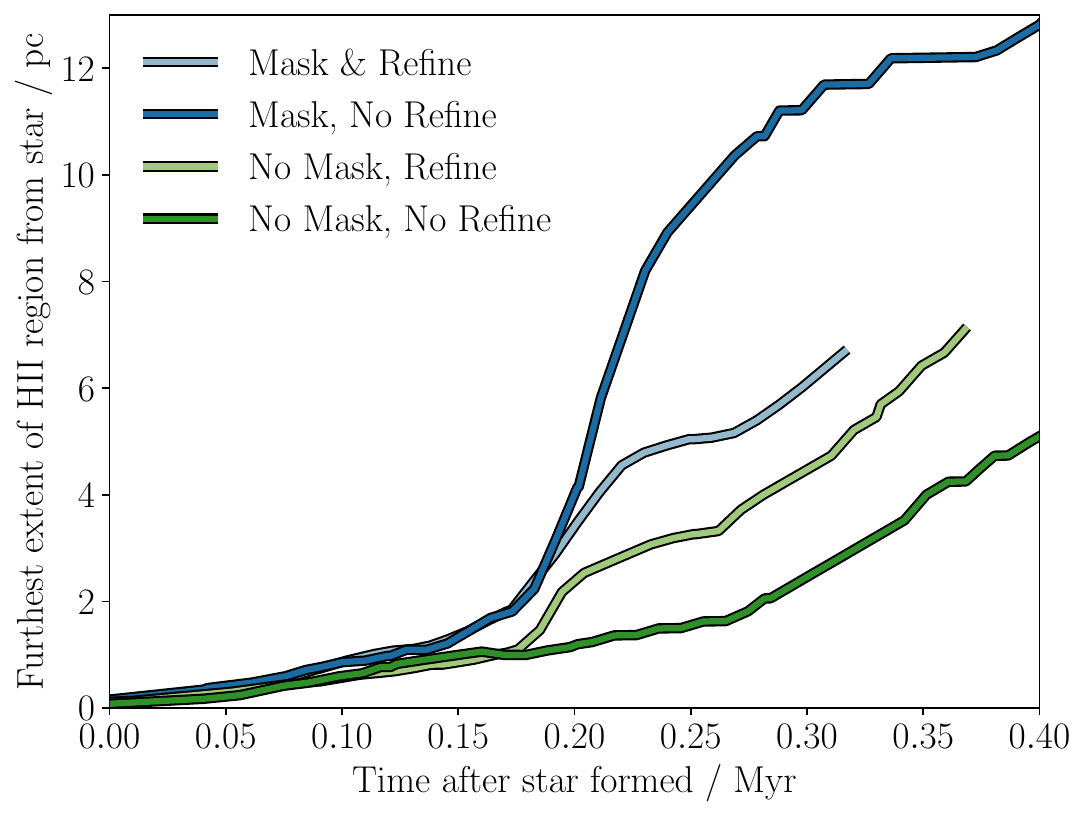}
\includegraphics[width=\columnwidth]{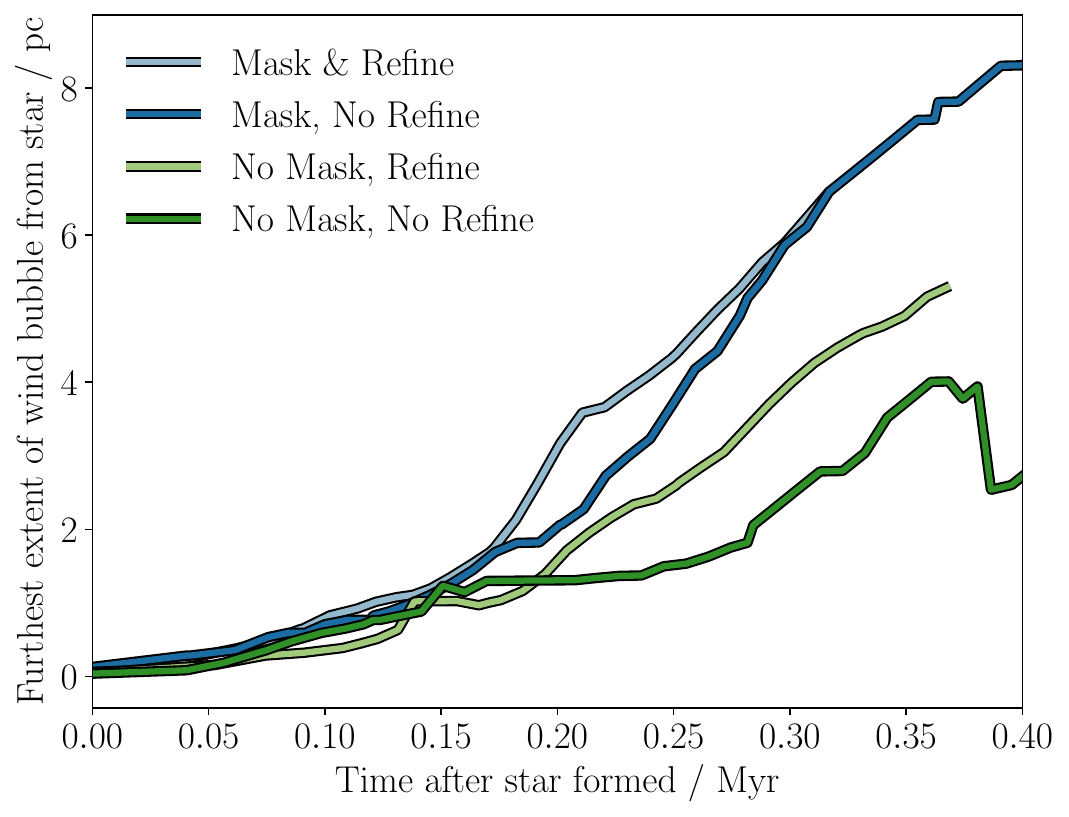}
\caption{Maximum extent away from the star of the \HII region (left) and the wind bubble $R_w$ (right) as a function of time in each simulation in the \textsc{physics} set. Winds act to constrain runaway photoionisation of the whole cloud by trapping the radiation in the shell around the wind bubble.}
\label{fig:radiusphysics}
\end{figure*}

\subsubsection{Wind-only Comparison}
\label{results:windonlycomparison}

We first compare results in a test setup without ionising radiation from the star, in order to simplify the comparison between runs with and without the wind contact discontinuity cooling mask. 

In Figure \ref{fig:windonlytest} we compare the radial extent of the wind bubble as a function of time and the energy from stellar winds retained in the wind bubble with and without the cooling mask. The spherical-equivalent radius (calculated assuming the volume of hot gas $V_{w}$ above $10^5~$K is a sphere, i.e. $(3V_{w}/4\pi)^{1/3}$) at 300 kyr is typically half as large when the cooling mask is applied versus without, although the stellar wind energy retained is higher when the cooling mask is applied. This slower radial expansion rate is somewhat counterintuitive, as one might expect more retained energy in the wind bubble to equate to a faster expansion rate. Indeed, this effect is reversed when we include photoionisation. We discuss the reasons for this below. Similarly, analysing the simulations further, the momentum added to the cloud by the wind is roughly doubled when the cooling mask is applied, although the pressure in both regions is very similar. 

Another major difference between the behaviour of the wind bubbles with and without the cooling mask is seen by analysing a slice through the bubble in Figure \ref{fig:windonlyslices}. Though certain parts of the wind bubble have extended further when the cooling mask is applied, there is considerably more substructure inside the bubble itself, seen as the yellow ridge inside the red bubble. This in turn reduces the spherical-equivalent mean radius, which does not include any dense, neutral substructures inside the wind bubble, only cells heated to above $10^5~$K.

As seen in Figure \ref{fig:windonlyslices}, runs where we suppress cooling at the contact discontinuity exhibit increased macroscopic mixing with the cloud material around the wind bubble. This is likely because energy losses via the turbulent cascade into thermal energy are suppressed. However, environment around the star is turbulent with large density gradients, complicating a simple linear analysis, and it is equally possible that a small change in the simulation leads to diverging behaviour as the turbulent system evolves, and differences over time become the result of non-linearity inherent in the fluid equations. Overall, applying the cooling mask thus does not prevent the wind bubble cooling entirely -- rather, it sets a lower bound on unresolved ``sub-grid'' cooling, where not applying the mask gives an upper bound that assumes that any unphysical numerical cooling is equivalent to that from any unresolved mixing or conduction processes. 

To summarise, the cooling mask does increase the amount of energy retained in the wind bubble. However, due to interactions with the turbulent cloud, this does not necessarily lead to a faster radial expansion of the wind bubble, and the mask instead enhances the role of macroscopic turbulent mixing on the scale of a few cells as opposed to sub-grid mixing, which is suppressed by the mask. The downside is that the mask suppresses both physical, unresolved cooling \citep{Rosen2014,Lancaster2021a} and non-physical numerical overcooling \citep{Fierlinger2016,Gentry2017}, although this is unavoidable without fully resolving the wind bubble contact discontinuity, which in turn becomes computationally expensive.

\subsubsection{Comparison including Photoionisation and Refinement}

We now consider a comparison of the effect of the cooling mask including photoionisation feedback and refinement on pressure gradients, i.e. at the edge of the \HII region. In Figure \ref{fig:energyretainedphysicsfb} we plot the energy retained in the wind bubble with and without the cooling mask and refinement on pressure gradients (left) as well as with and without photoionisation (right). We define the wind bubble as including all cells with a temperature above $10^5~$K, where energy is the total thermal and kinetic energy.

Unsurprisingly, the cooling mask causes the wind bubble to retain more energy, although the difference shrinks if refinement on pressure gradients is included, suggesting that the results should converge if increased resolution across the contact discontinuity is achieved (with the corresponding increase in computational requirements).

The inclusion of photoionisation also increases the energy retained in the wind bubble. Photoionisation heats the gas around the wind bubble to $\sim 10^4~$K, which both reduces the temperature difference between this gas and the hot wind bubble and causes it to thermally expand to a lower density. This means the wind bubble mixes with a relatively warmer, more diffuse medium than the cold, dense neutral shell, leading to reduced cooling in the wind bubble.

We plot slices through the wind bubble, photoionised \HII region and surrounding cold cloud in Figure \ref{fig:imagephysics}. The wind bubble is highly aspherical, forming a chimney of hot, fast-flowing gas that pushes through low-density channels in the cloud. This chimney becomes a wider plume of hot gas as it encounters the interface between the \HII region and the unshocked gas outside. As refinement is turned on, this plume becomes more structured with fluid instabilities, in some cases breaking up into isolated hot bubbles. Warm photoionised gas fills in the volume between the star and the edge of the \HII region. In the ``Mask, Refine'' simulation, the photoionised gas does not extend further than the wind bubble, although in all cases a full champagne flow is prevented.

We study the radial evolution of the wind bubble and \HII region in more detail in Figure \ref{fig:radiusphysics}. We plot here the maximum radius, i.e. the distance of the furthest cell in the wind bubble or \HII region from the star. Measuring the radius by using the \HII region volume $V_{HII}$, where $R_{HII} = (3V_{HII}/4\pi)^{1/3}$, gives a similar result. However, in certain simulations, e.g. ``No Mask, No Refine'', the wind bubble is less extended but has a larger volume. We use this comparison here since the wind bubble is highly aspherical and its most observable characteristic is its extent.

Masking cooling at the contact discontinuity allows for a more extended \HII region than for runs without the mask. Unlike the Wind Only simulations, UV radiation photoionises any substructure trapped by the wind bubble. As the \HII region expands further, the wind bubble also expands into the underdensity created by the \HII region, expanding further in runs with the cooling mask applied. Increased refinement either gives similar or a larger radius. However, in the case of the \HII region radius, the ``Mask, No Refine'' run reaches the edge of the cloud and begins to break out in one direction, despite having a similar wind bubble extent to the run with refinement.

The complex, non-linear gas dynamics displayed makes it hard to demonstrate a consistent linear response to the different physical models used. However, all models display a similar behaviour, with increased refinement leading to the wind bubble displaying more turbulent substructures. Including a cooling mask also leads to a generally wider plume as well as a longer chimney.

\subsubsection{The Role of Turbulence}

\begin{figure*}
	\centering
	\includegraphics[width=0.45\hsize]{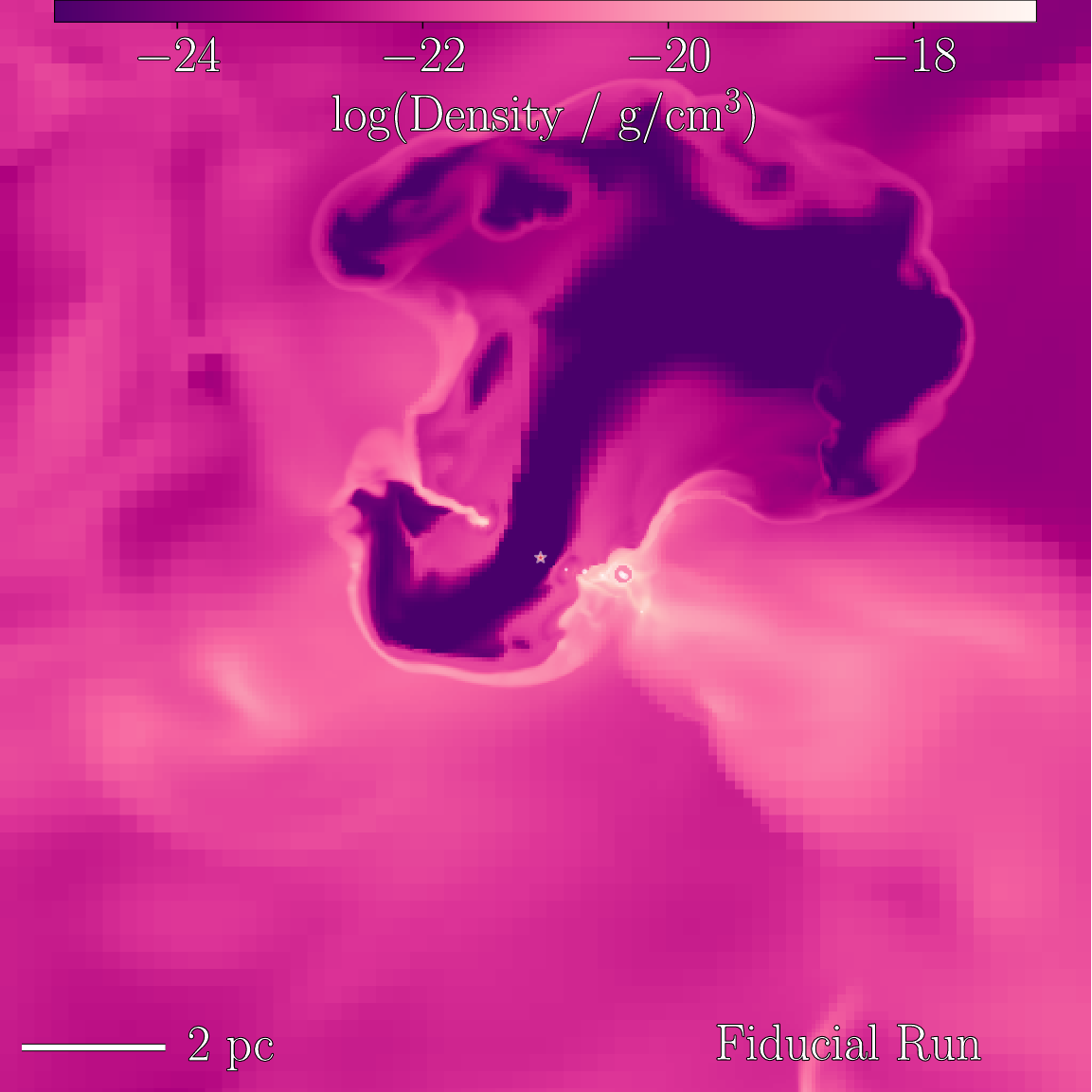}
	\includegraphics[width=0.45\hsize]{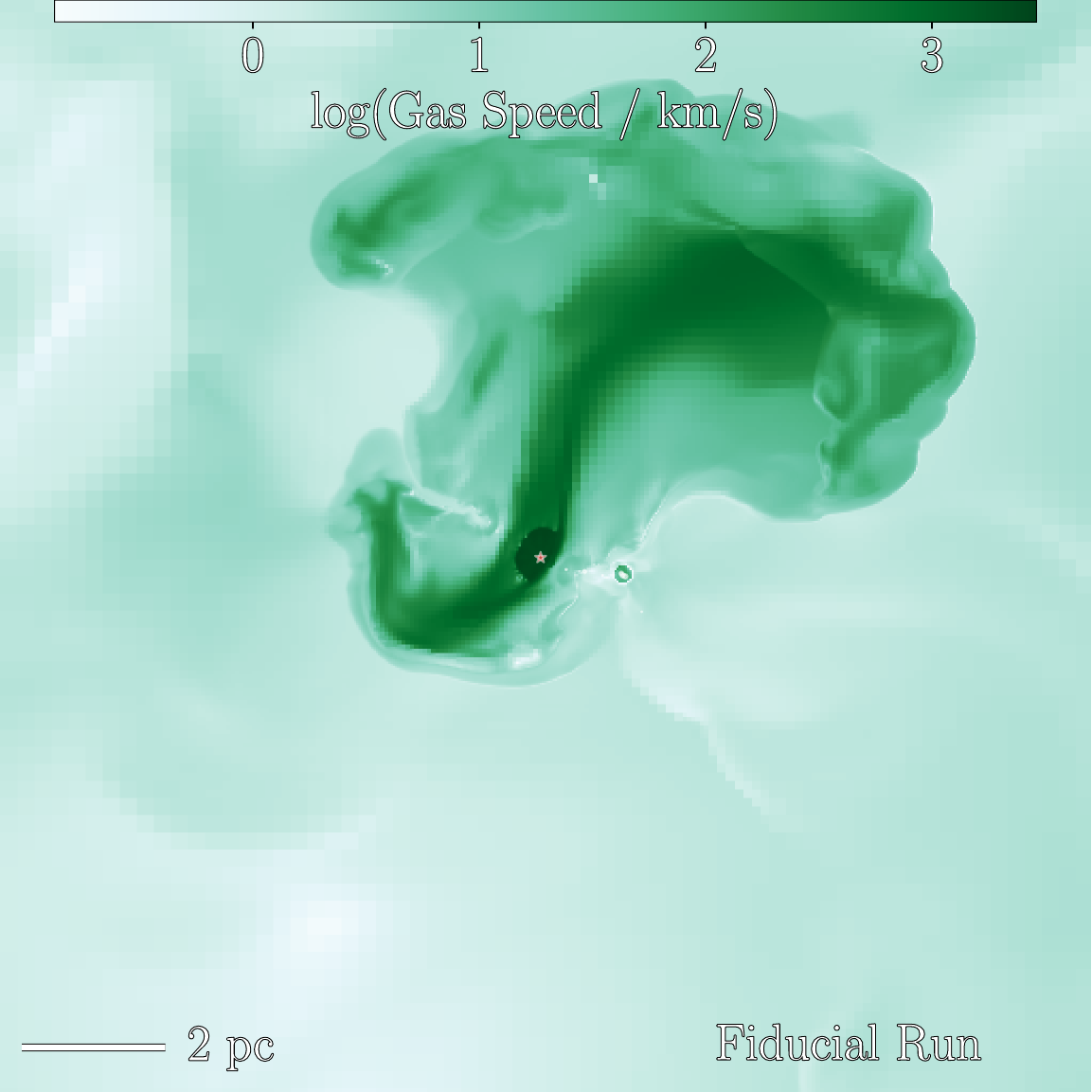}
	\includegraphics[width=0.45\hsize]{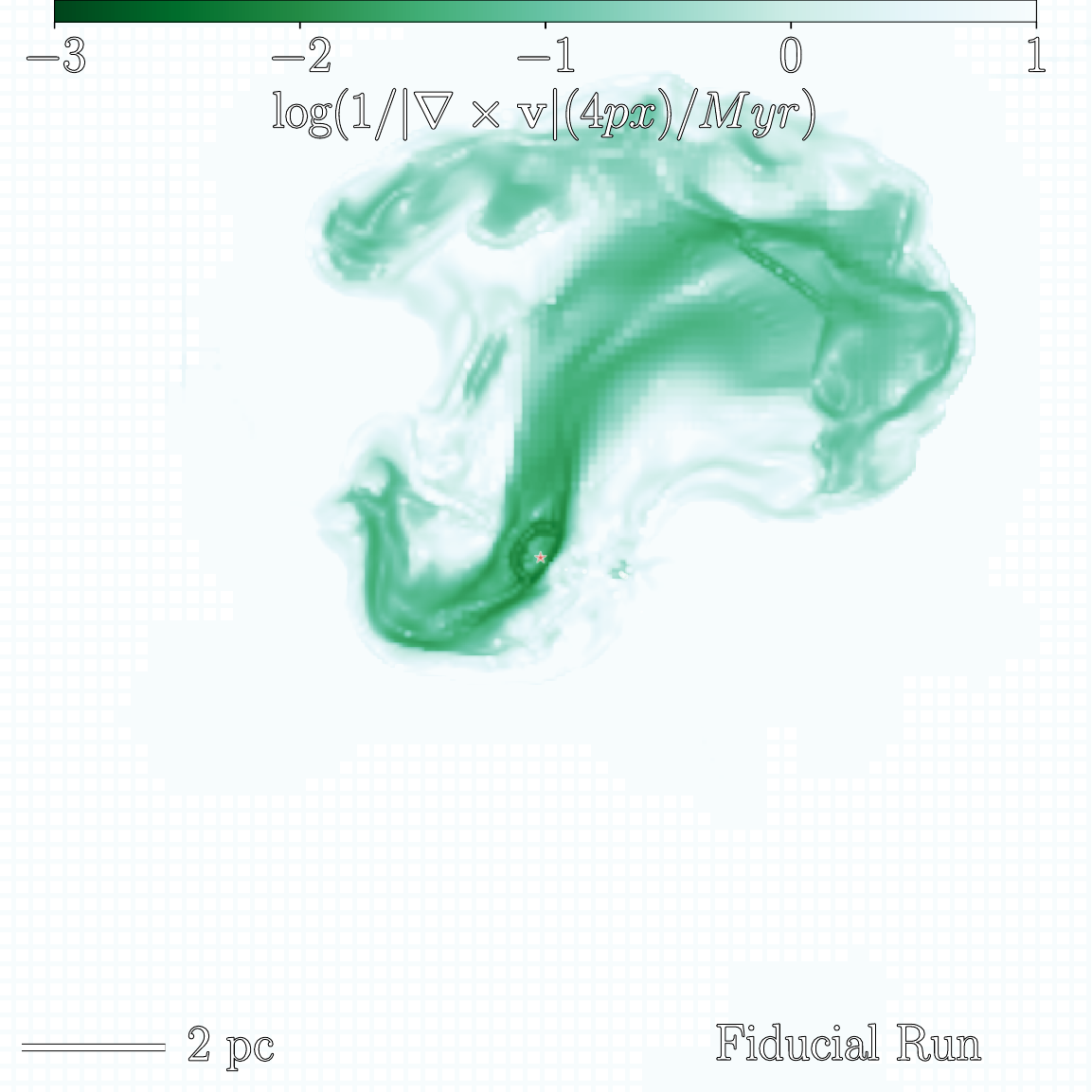}
	\includegraphics[width=0.45\hsize]{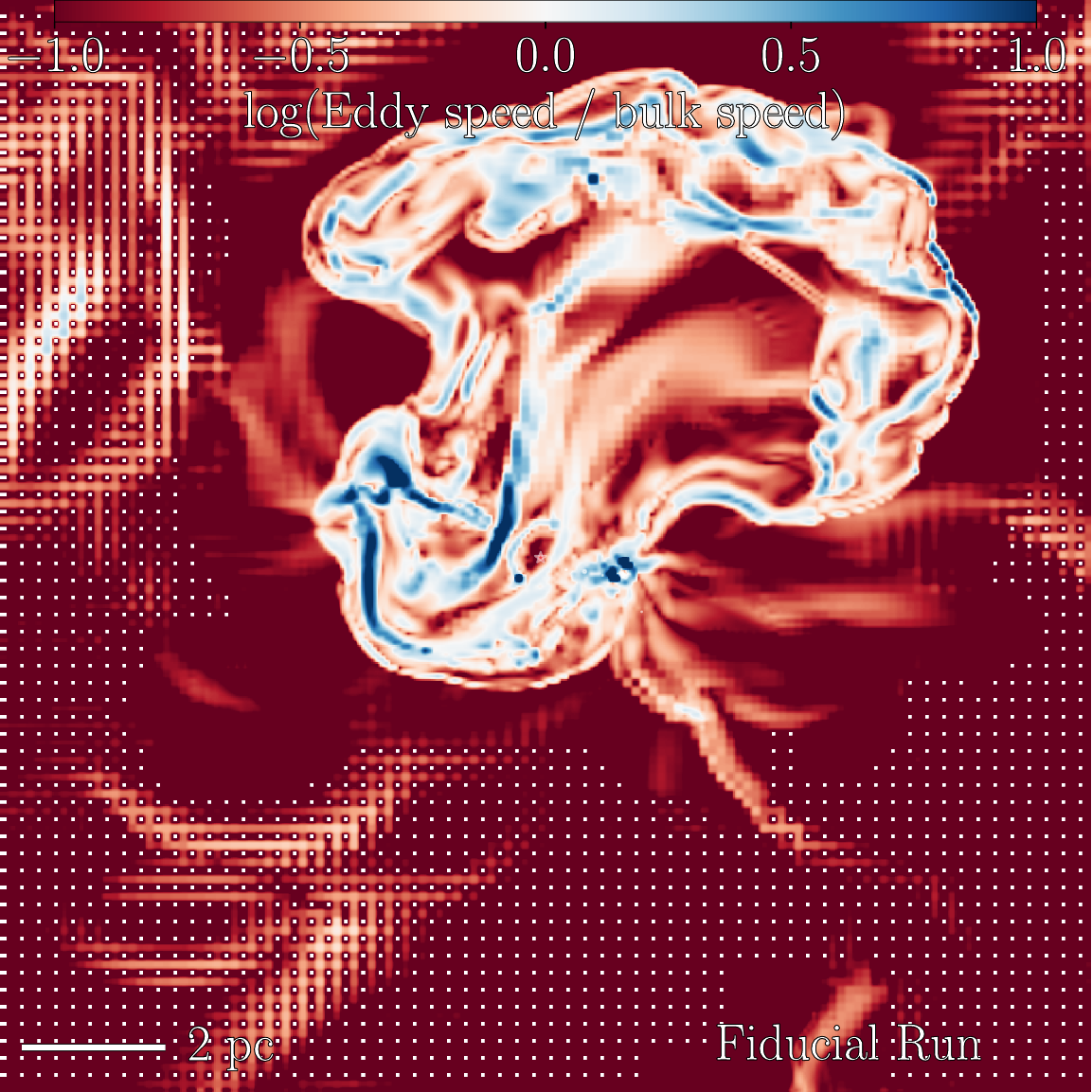}
	\caption{Slices through the Fiducial Run at a stellar age of 0.3 Myr, depicting the turbulent properties of the gas. From top left to bottom right, we show gas density, bulk gas speed, the inverse of gas vorticity (i.e. the eddy turnover timescale at a scale of 0.12 pc (4 times the smallest cell size)), and the ratio of characteristic eddy speed ($\ell \nabla \times \mathbf{v}$) to bulk gas speed. By comparing the eddy speed to the bulk gas speed we aim to identify where the gas flows are mostly laminar at a 0.12 pc scale, and which flows are highly turbulent. The interior of the wind bubble is indeed highly turbulent, with mixing evident between the wind bubble and the rest of the \HII region. The circle around the star is the free-streaming radius of the wind. The white dots in the bottom right figure are where the lower grid resolution creates points where a cell is compared to itself, resulting in zero eddy speed.}
	\label{fig:turbulenceslices}
\end{figure*}

\begin{figure}
	\centering
	\includegraphics[width=\columnwidth]{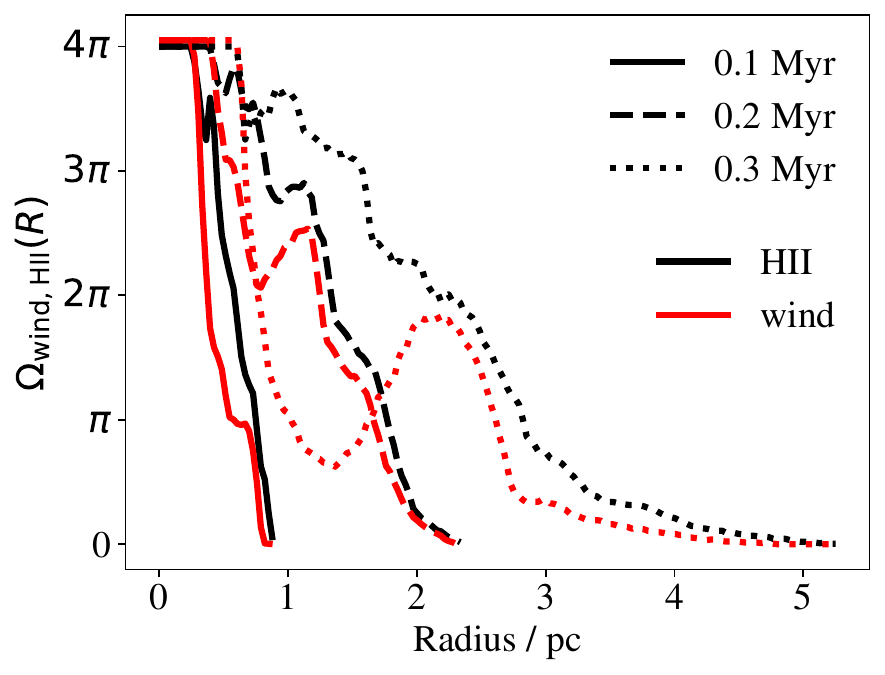}
	\caption{Solid angle subtended by the wind bubble (red) and the overall \HII region (black, includes both photoionised and wind-shocked gas) as a function of radius in the Fiducial Run. Results taken at stellar ages of 0.1 Myr are shown as a solid line, 0.2 Myr as a dashed line and at 0.3 Myr as a dotted line. We project 1000 rays away from the star, and measure the fraction of the full solid angle around the star ($4 \pi$) that contains either wind-shocked gas ($T > 10^5~$K) or photoionised gas ($x_{HII} > 0.1$) at each radius from the star. Over time, a wide plume develops that has a larger solid angle than the chimney that feeds it. In many cases, photoionised gas back-fills the solid angles subtended by the plume at smaller radii, since the ionising radiation travels directly outwards rather than seeking pressure gradients as the wind bubble does.}
	\label{fig:solidangle}
\end{figure}

In the classical \cite{Weaver1977} model of wind bubbles, the majority of cooling takes place via evaporation of material from the dense shell into the hot wind bubble \citep{MacLow1988}, which is slow even if the wind bubble is embedded in a dense molecular cloud environment \citep{Kruijssen2019,Geen2022}. However, wind bubbles evolve into complex structured environments, and develop fluid instabilities that are impossible to simulate directly in simple 1D models, and which can enhance cooling via turbulent mixing \citep{Rosen2014}. \cite{Lancaster2021a} present a model of the evolution of wind bubbles with a fractal surface that assumes efficient cooling via turbulent mixing with the medium around the wind bubble. The authors describe a number of diagnostics to track where this efficient cooling should occur, assuming a globally uniform cloud density field. This compares well to simulations of wind bubbles representing stellar populations in molecular clouds with idealised fixed sources \citep{Lancaster2021b} and with sources forming on sink particles \citep{Lancaster2021c}.

In our work, we focus on a single stellar source (versus multiple sources in \citealp{Lancaster2021c}) but also include a cooling mask designed to remove spurious numerical cooling. However, while this mask does remove artificial cooling, it also prevents all cooling at the wind bubble's contact discontinuity on the scale of 1-2 grid cells, including possible cooling from unresolved turbulent mixing. While \cite{Tan2021} argue that the largest eddy scale dominates turbulent mixing, \cite{Lancaster2021b} find that in simulations of wind bubbles in molecular clouds, the relevant cooling scale is difficult to resolve. Nonetheless, they argue that it is possible to determine if efficient turbulent mixing can occur if the turbulent velocity of gas at the wind bubble interface allows faster diffusion of energy than can be injected by the bulk velocity of material from the star entering this interface.

Turbulence is a chaotic phenomenon characterised by fluid eddies at multiple scales which diffuse energy from large to small scales. There are a number of ways to characterise this in simulations. The most simple method for cases where there is one source is to assume that the source will generally produce radially-aligned outflows, and thus measuring non-radial flows gives an estimate of the levels of turbulence in the outflows as opposed to laminar radial flows. However, this is not possible in our case since even the broadly laminar chimneys flowing from the star become non-radial due to following decreasing density contours in the cloud. Another is to measure the velocity dispersion, which is a statistical measure of how different fluid velocities are in surrounding computational elements. Finally, one can measure the vorticity, i.e. curl of the velocity field ($\nabla \times \mathbf{v}$), which gives the turnover frequency of fluid eddies at a given scale.

In Figure \ref{fig:turbulenceslices} we show slices through the wind bubble in the Fiducial Run to analyse the impact of turbulence on the wind bubble using vorticity as the measure of turbulence. We sample across a nominal length scale $\ell$ of 4 simulation cells at the highest level of refinement ($\ell=0.12~$pc) in order to smooth out noise on a cell-by-cell level. In this figure we plot the gas density, the bulk gas speed, the eddy turnover timescale ($1 / \nabla \times \mathbf{v}$) and the ratio between the characteristic eddy speed $\ell \nabla \times \mathbf{v}$ and the bulk gas speed in each cell. Some grid artefacts can be seen due to the lower cell resolution outside the wind bubble where we do not fully refine the grid. 

The eddy turnover timescale is longest (i.e. darker green in the bottom left panel of Figure \ref{fig:turbulenceslices}) in the chimney of the wind bubble, as well as the extended plume at the edge of the bubble, while the bulk gas velocity is highest in the chimney. There are regions with strong eddies on the shearing interface between the chimney and the photoionised gas outside, as well as on the complex surface where the plume interacts with the surrounding material. There are also strong eddies in the dense shell itself, although this can also be a measure of the velocity difference between the expanding wind bubble and the undisturbed material outside it.

The bulk speed of the gas in the wind bubble, and the expansion rate of the wind bubble itself, is significantly higher than the speed of the gas outside the wind bubble. This suggests that much of the turbulence at the wind bubble interface is shear from fast-moving chimneys of gas expanding into regions of lower density. 

Ultimately, the ``correct'' modelling of the effect of turbulence on the scale of a few cells around the wind bubble interface remains a difficult task, and the true fraction of energy lost to cooling will lie somewhere between the models with and without the cooling mask. The ``true'' cooling rates and hence efficiency of wind bubbles as feedback sources thus remains a subject of difficulty, despite promising work in recent years. Some of these difficulties can be mitigated by using Lagrangian (i.e. moving mesh) methods versus static Eulerian grids. However, spatial resolution requirements remains a tough limitation to overcome, particularly given the already difficult computational requirements in capturing the timesteps needed to trace flows travelling at 1000s of km/s.

\subsubsection{How do physical wind bubbles expand rapidly?}

Analysis of the Orion Nebula M42 by \cite{Pabst2019} finds that the wind bubble expands at a speed consistent with an adiabatic wind bubble, or $\sim13~$km/s. Conversely, following our simulations, we do not expect the wind bubble interface to be adiabatic as the wind bubble interface exhibits large quantities of turbulence on all scales, which we expect to lead to enhanced cooling. This in turn should lead the wind bubble to expand more slowly, as it has a lower amount of energy available to drive its expansion.

A cooling mask does help to mitigate this by removing sources of overcooling from artificial numerical mixing. However, it does not completely remove sources of cooling to an adiabatic level. When comparing simulations with and without a cooling mask in Figure \ref{fig:energyretainedphysicsfb}, we find a factor of a few difference in cooling when applying the cooling mask versus not in cases where we also include refinement on pressure gradients. Applying the cooling mask in these cases allows the wind bubble to retain $\sim5-7~$\% of the energy injected by the star in the gas (including both thermal and kinetic components). This is compared to 1-2$~$\% when no mask is implied. This is significantly lower energy retention than the 33\% of wind energy retained (after work done on the surrounding medium is included) found in the adiabatic model of \cite{Geen2022} assuming a singular isothermal sphere power law density field, or 45\% retention when using the \cite{Weaver1977} model in a uniform medium. 

One explanation is that the opening angle of the wind chimney is significantly smaller than the total opening angle of the wind bubble. This creates a wind bubble that has a smaller volume than a spherical wind bubble with the same extent away from the star, and may explain why the wind bubble is able to reach larger radii. We plot this in Figure \ref{fig:solidangle}, showing the solid angle subtended by the wind bubble (in red) and the total \HII region (in black) at each radius away from the star for a set of stellar ages. As the wind bubble evolves, the region around the star remains fully wind-driven (partly caused by the free-streaming phase of the wind), before forming chimneys that take up under $\pi$ of the full $4 \pi$ of the solid angle. This is an average across all directions from the star including the smaller "back-blast" seen in the bottom left of each panel in Figure \ref{fig:turbulenceslices} and earlier figures showing the same simulation. 

Additionally, by examining both Figures \ref{fig:imagephysics} and  \ref{fig:turbulenceslices}, we see that the picture in our 3D turbulent cloud is very different to that in spherically symmetric 1D models where a spherical wind bubble sits inside a photoionised shell. While the hot wind bubble follows regions of low density, radiation from the star must travel in straight lines, and so fills a larger solid angle around the star than the wind bubble, which travels in a narrow chimney through the cloud before expanding into a wide plume. This implies that some \HII regions appearing to have large wind bubble volume filling fractions may in fact exhibit a large plume near the edge of the \HII region with a narrow embedded chimney. It also suggests that the wind bubble may expand faster than in a purely spherical model if it is allowed to expand over narrow solid angles.


\subsection{The Role of Chaos and Initial Conditions in Shaping Wind Bubbles}
\label{results:seeds}

\begin{figure*}
	\centering
	\includegraphics[width=\hsize]{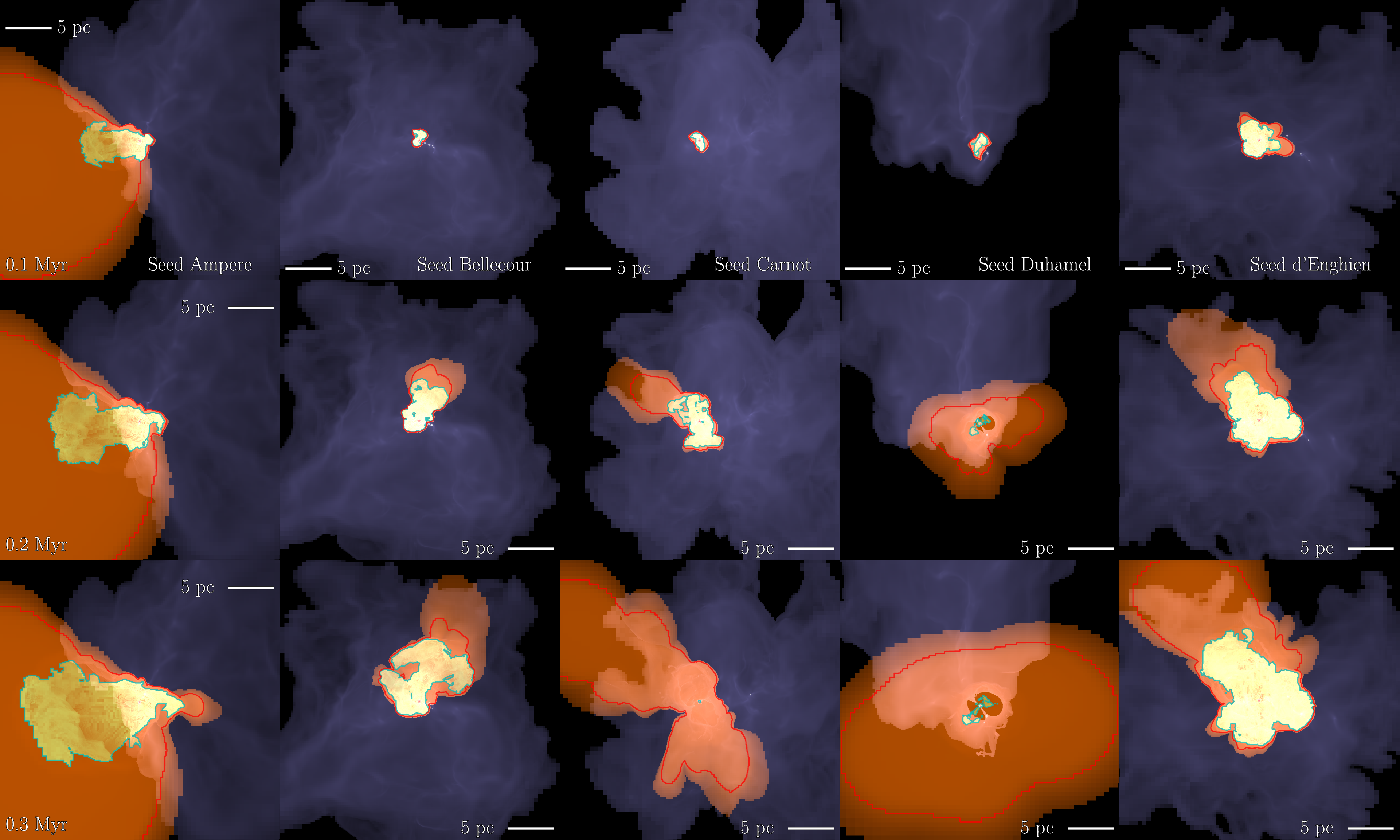}
	\caption{As for the Wind \& UV simulation shown in Figure \ref{fig:imagefb} but with the \textsc{seeds} set showing the effect of different random seeds in the initial turbulent field on the resulting nebula.}
	\label{fig:imageseeds}
\end{figure*}

\begin{figure}
	\centering
	\includegraphics[width=\columnwidth]{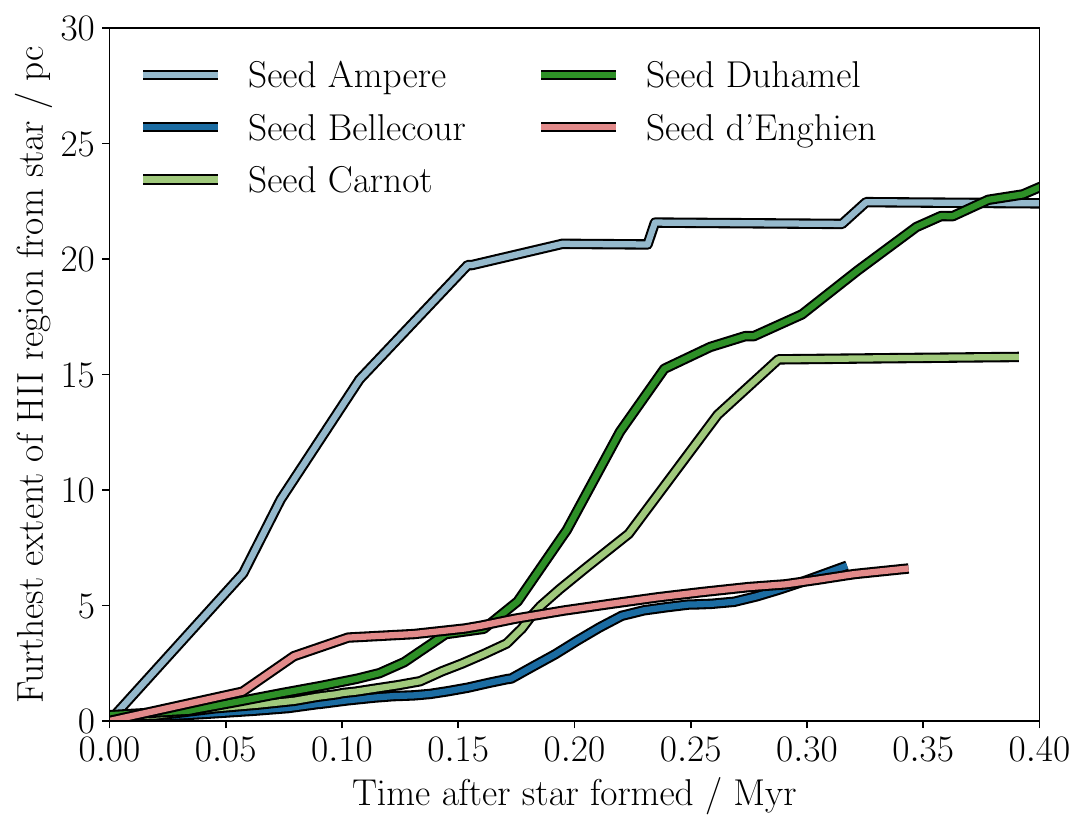}
	\caption{Maximum extent of the \HII region away from the star (including wind and photoionisation-heated gas) as a function of time in each simulation in the \textsc{seeds} set. Winds act to constrain runaway photoionisation of the whole cloud by trapping the vast majority of the radiation in the shell around the low-density wind bubble, as described in Section \ref{results:combiningwindsandradiativefeedback}. Fast radial expansion in the \textsc{Ampere}, \textsc{Carnot} and \textsc{Duhamel} runs is caused by the \HII region encountering the edge of the cloud and expanding into the low-density medium outside as a ``champagne'' flow.}
	\label{fig:radiusseeds}
\end{figure}

\begin{figure}
	\centering
	\includegraphics[width=\columnwidth]{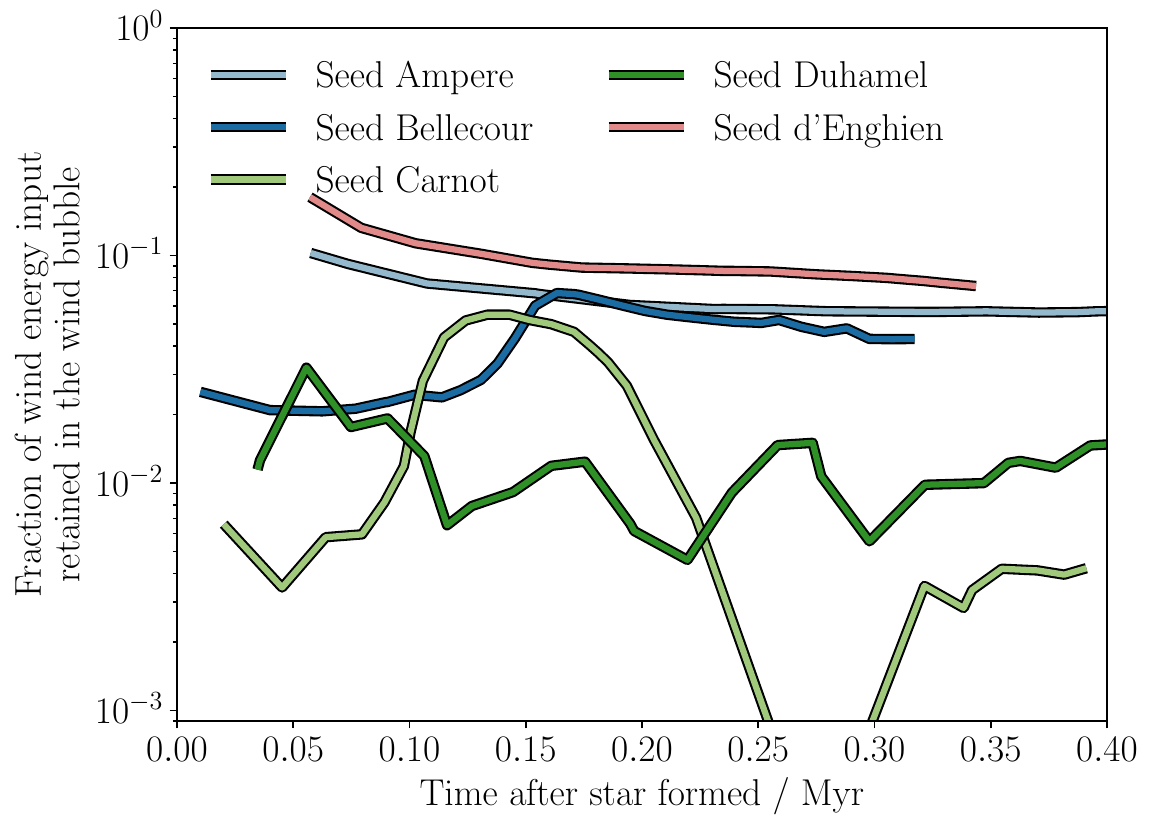}
	\caption{The energy in the wind bubble as a fraction of the total energy injected as stellar winds, as a function of time in each simulation in the \textsc{seeds} set. Strong wind bubbles retain up to 10\% of the energy from stellar winds, while weak bubbles retain around 1\%, with some results moving between the two limits. (For reference, an adiabatic wind bubble evolving in a singular isothermal sphere retains 33\% of its energy inside the bubble, see \citealp{Geen2022}). The loss of the wind bubble in the Seed \textsc{Carnot} run due to the emergence of a ``hot champagne'' flow is visible as the sharp decline around 200 kyr. }
	\label{fig:energyretainedseeds}
\end{figure}

\begin{figure*}
	\centering
	\includegraphics[width=0.3\hsize]{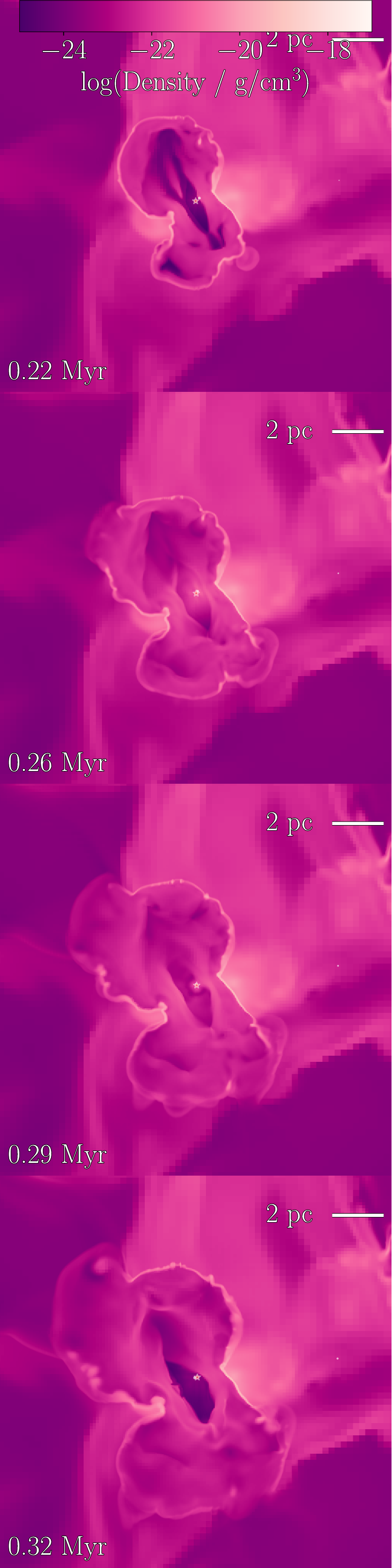}
	\includegraphics[width=0.3\hsize]{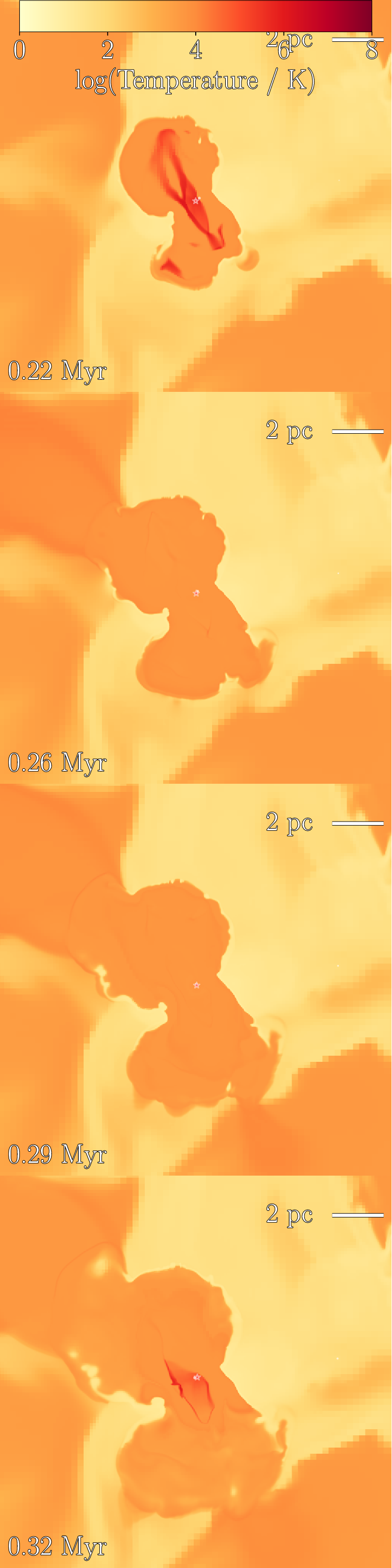}
	\includegraphics[width=0.304\hsize]{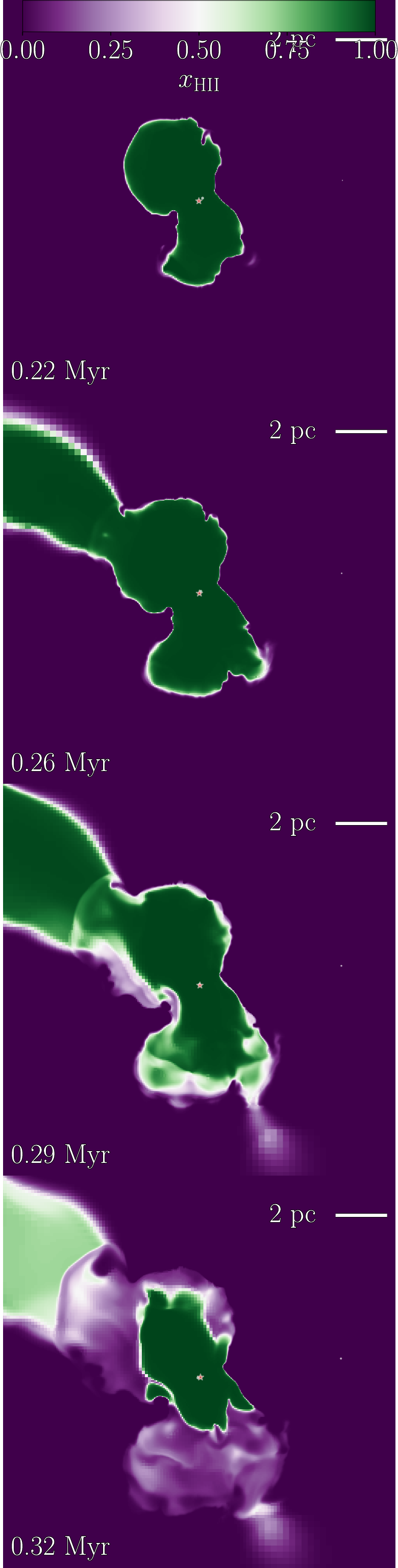}
	\caption{Slices through the star in the Seed \textsc{Carnot} run, depicting the ``hot champagne'' outflow and disappearance of the wind bubble. Fields from left to right show gas density, temperature and photoionised fraction (ignoring collisional ionisation, which occurs at higher temperatures in the wind bubble). Time evolves from top to bottom, showing snapshots as photoionised gas breaks out into a champagne flow, where the wind bubble mixes efficiently with the photoionised gas to the point where no hot ($>10^6~$K) gas remains visible. The phenomenon is transient and lasts for less than $0.1~$Myr.}
	\label{fig:hotchampagne}
\end{figure*}

\begin{figure*}
	\centering
	\includegraphics[width=\hsize]{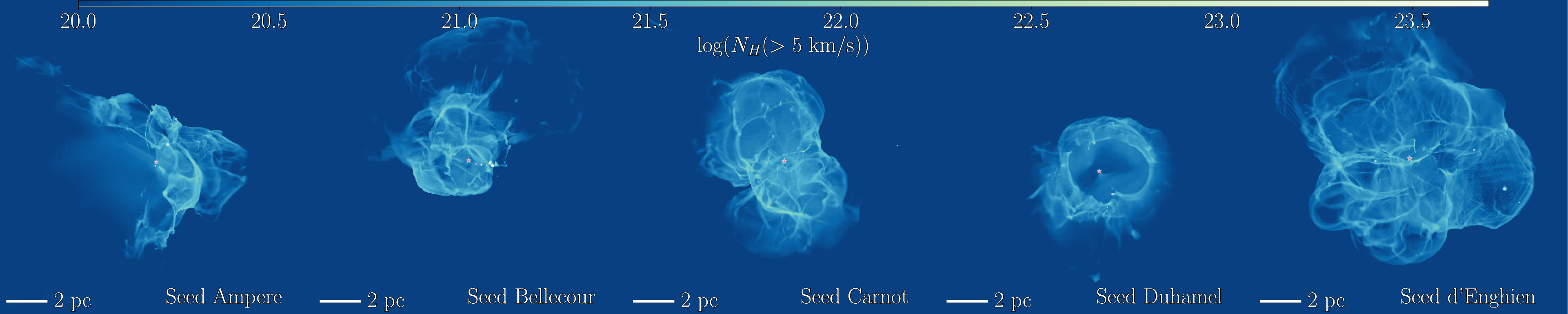}
	\includegraphics[width=\hsize]{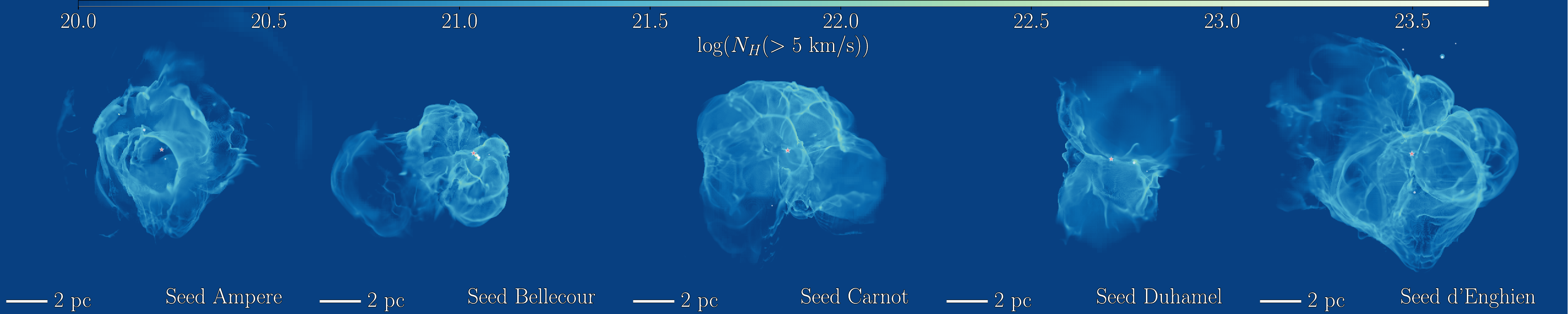}
	\caption{Projections of column density of gas cells with a speed greater than 5 km/s in the \textsc{seeds} set. The top row is projected along the x-axis, and the bottom row along the z-axis. The goal is to track mass swept up by \rev{feedback} and identify where dense shell-like structures remain. Photoionised champagne flows are evident where dense bubble-like structures have been burst, e.g. in the top left panel, where the expanding gas on the left side of the nebula merges into the background without a clear transition indicating a dense shell.}
	\label{fig:fastmass}
\end{figure*}

\cite{Geen2018} demonstrate how the role of chaos plays a significant role in the outcomes of feedback in star-forming clouds. The use of different random seeds to initialise the turbulent velocity field in the cloud, which in turn sets the distribution of density peaks that form in the cloud, causes feedback to behave in different ways.

In Figure \ref{fig:imageseeds}, we plot the results of the \textsc{seeds} simulation set where we vary the random seed \rev{used to generate the initial conditions, as described in Section \ref{methods:initial_conditions}}, and explore what effect this has on the resulting wind bubble and \HII region. We form the stellar source self-consistently following sink particles, and use the same stellar tracks (a 35 \Msolar star) in each simulation. 

There is a large diversity of behaviours of the \HII regions purely by varying the initial seed. In some cases the photoionised region is effectively contained near the wind bubble, even after a few hundred kyr, but in other cases there is still a rapid champagne flow out of the cloud. Similarly, some simulations have a prominent wind bubble, while others have a small wind bubble. The wind bubble in Seed \textsc{Carnot} even grows and disappears, before appearing again 0.1 Myr later, which we discuss below.

The emergence of a champagne flow breakout or otherwise in the \HII region for each seed is visible in Figure \ref{fig:radiusseeds}, which is displayed as a sudden increase in radial expansion rate much faster than the gas sound speed of $\sim$10 km/s $\simeq$ 1 pc / 0.1 Myr. Seeds \textsc{Bellecour} and \textsc{d'Enghien} never display a strong champagne breakout. By comparison, in Seed \textsc{Ampere}, the seed used in \cite{Geen2020}, the breakout begins almost immediately (see also the top row of Figure$~$\ref{fig:imageseeds}).

A key aspect of whether a champagne flow forms is whether the star forms on the edge of the cloud or not. The trapping of ionising radiation by the wind bubble described in \cite{Geen2022} requires a supply of swept-up cloud material to maintain the trapping. If the wind bubble encounters a density drop at the edge of the cloud, the shell can ```burst'', leading to a champagne flow as described in \cite{TenorioTagle1979} and \cite{Whitworth1979} \citep[see also][for models including winds]{Comeron1997}.

In Figure \ref{fig:energyretainedseeds} we plot the energy emitted by the star as stellar winds retained in the wind bubble. An adiabatic wind should retain 45\% of its energy in the hot wind bubble for a uniform cloud \citep{Weaver1977} or 33\% in a $\rho \propto r^{-2}$ power law density field \citep{Geen2022}. Three of the seeds retain 10-20\% of their energy at some point in their evolution, which approaches adiabatic assuming the power law density field around the star, with the retained energy slowly decreasing over time. We discuss the cases of Seeds \textsc{Carnot} and \textsc{Duhamel} in the following two sections, and discuss their behaviour. Seed \textsc{Carnot} undergoes a hydrodynamically unstable process, which drastically increases mixing of the wind bubble. In the case of Seed \textsc{Duhamel}, the low energy retention is likely to be due to resolution limits rather than a physical lack of energy retention. Nonetheless, despite the differences in the wind bubble evolution, 3 out of 5 of the simulations display evolution in their energetics after the first 0.1 Myr.

\subsubsection{``Hot'' Champagne}

Seeds \textsc{Ampere}, \textsc{Bellecour} and \textsc{d'Enghien} display large-developed wind bubbles. By contrast, Seed \textsc{Carnot} loses its wind bubble around 0.3 Myr, while Seed \textsc{Duhamel} never develops a strong wind bubble at all. This is not a function of strong accretion onto the sink in these cases - although Seed \textsc{Duhamel}'s star does continue to be embedded inside a large filament up to 0.3 Myr, accretion rates are similar in all runs, except for Seed \textsc{Bellecour} which is higher despite having a well-developed wind bubble. Similarly, the ram pressure of flows in the neutral gas around the sinks at the time of formation is similar.

In the case of Seed \textsc{Carnot}, the disappearance of the wind bubble is linked to the \HII regions entering a breakout champagne flow phase. Figure \ref{fig:hotchampagne} displays the progression of the \HII region at the point a champagne flow begins. The depressurisation of the photoionised region allows the wind bubble to expand more quickly and mix with the photoionsed gas, causing a slightly hotter \HII region with the same temperature as the now cooled wind bubble. As the wind is injected on top of the background gas, turbulent mixing near the star can cause the wind bubble to disappear temporarily even if the wind energy is still being injected.

We refer to this as a ``hot champagne'' flow, since it involves the efficient mixing of hot ($>10^6~$K) wind-shocked and warm ($\sim 10^4~$K) photoionised gas. The phenomenon is transient and lasts for less than $0.1~$Myr. Once the initial rapid expansion phase has occurred, the wind bubble gradually reappears (see the lower panel of Figure \ref{fig:hotchampagne}. This is combined with other effects such as photoionisation in the champagne flow being temporarily disrupted. As the phenomenon is linked to out-of-dynamical-equilibrium behaviour in the \HII region following its encountering a discontinuity, this effect is also temporary.

If found in physical \HII regions that have just begun to undergo the champagne phase, this behaviour is likely to pose a problem for observational studies that map the temperature of the photoionised region to the spectrum of the star, assuming that residual energy from the wind bubble remains in the photoionised gas for an appreciable amount of time. A \HII region without a visible wind bubble may be an indication that such a phenomenon has occurred. More simulation work is needed to establish how common this phenomenon is, as well as whether it can be found in other cosmic conditions such as at lower metallicities.

\subsubsection{Shell Structure}

Comparisons with full synthetic observations are left to future work, given the complexity of matching the full velocity structure of line emission from tracers such as C$^{+}$ \citep[e.g.][]{Schneider2020}. However, we approximate the observable features of the neutral shell around the \HII regions by measuring the column density of gas in cells with a total speed faster than 5 km/s, in order to isolate material in the dense, expanding shell. This is also where tracers such as C$^{+}$ would be found, while also removing ambient gas in the cloud that is typically moving at lower velocities. We plot this quantity in Figure \ref{fig:fastmass}. 

In the top image (x-projection, used in the other figures in this paper), large-scale photoionised champagne flows, evidenced by the lack of a shell over a large solid angle, are visible in Seed \textsc{Ampere}, but less so in other images. By contrast, in the bottom image (z-projection), the champagne flow is more visible in seeds \textsc{Duhamel} and \textsc{d'Enghien}. Analysis of the full velocity cube may be necessary to find gaps in the shell.

While we do not perform a direct comparison here, the multiple bubbles observed by \cite{Beuther2022} in NGC7538, powered primarly by an O3 star, are recovered. This is caused by preferential expansion in multiple directions by the wind bubble creating chimneys following density minima in the cloud, followed by spreading out in a plume outside the minima. This is visible in a 2D slice in Figure \ref{fig:hotchampagne}.

\subsubsection{The Role of Resolution}

A further point that will likely affect our results is the requirement for a minimum injection radius for stellar winds. \cite{Pittard2021} argue that in order for stellar winds to be produced, the wind must be injected within a radius of $r_{ini,max} = \sqrt{p_{wind} / (4 \pi P_{amb})}$, where $p_{wind}$ is the momentum of winds from the star and $P_{amb}$ is the ambient pressure. The reason for this is that at larger injection radii, the initial phase of overpressurisation in the wind bubble is not resolved, and thus the resulting bubble risks stalling or expanding too slowly. Our star has a momentum injection rate of $5.7\times10^{27}~$g$~$cm/s. In our Fiducial model, the typical ambient pressure is $10^{-9}~$erg/cm$^{-3}$. This gives $r_{ini,max} = 0.22~$pc. By comparison, our injection radius of 5 cells at the highest refinement level gives an injection radius $r_{inj}$ of $0.15~$pc. This is sufficient to produce a wind, however \cite{Pittard2021} argue that a ratio $r_{ini} / r_{ini,max} < 0.1$ is desirable to reproduce the correct wind bubble dynamics. 

By comparison, in Seed \textsc{Duhamel}, the ambient pressure is roughly $10\times$ higher, $r_{ini,max}=0.069~$pc, or 2-3 cells at our highest refinement level. This offers a plausible explanation for why a wind bubble struggles to form in this environment. 

The conditions during the ``hot champagne'' event in Seed \textsc{Carnot} should not in principle violate the pressure-resolution requirement given above, with the temporary disappearance of the wind bubble being due to a large growth in turbulent mixing in the \HII region, including the location of the source star. However, we caution that resolution limits in general may affect results in surprising ways, and this phenomenon may be affected by both numerical and physical conditions around the star.

Noting the already strenuous requirements on producing high-resolution simulations that include stellar winds, an outstanding problem remains in identifying methods for producing stellar wind bubbles accurately in complex, high-pressure environments.

\subsubsection{Additional Effects}

\rev{It is worth pointing out that even with multiple realisations of the simulated system, our work misses some effects that likely shape regions such as Orion. For example, protostellar jets have been argued to shape similar star-forming nebulae, both in observations \citep{Kavak2022} and simulations (\citealp{Grudic2022}, \citealp{Verliat2022} and \citealp{Guszejnov2022}). Similarly, we do not include cosmic rays (see review by \citealp{Girichidis2020}), although \cite{Morlino2021} argues that wind bubbles may not be significant sources of cosmic rays. We also do not simulate a wide range of molecular cloud densities, masses and metallicities as found across galaxies, nor do we allow multiple massive stars to form as feedback sources. Expanding this work thusly by expanding the parameter space of the problem, though likely costly in computational resources, will likely be necessary to expand our understanding of how massive stars transfer energy to the wider universe.}

\section{Conclusions}
\label{conclusions}

\begin{figure}
	\centering
	\includegraphics[width=\hsize]{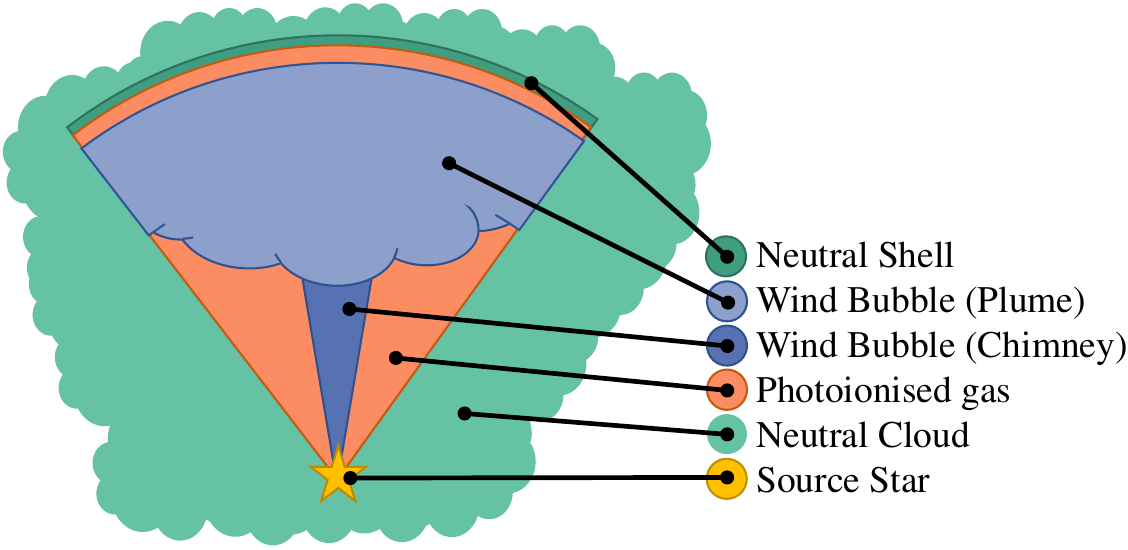}
	\caption{Diagram showing the schematic behaviour of an Orion (M42)-like nebula as modelled in our simulations. The winds stream away from the source star in a collimated chimney, constrained in other angles by denser gas around the star, before expanding into a turbulent plume. Photoionised gas fills out the volume up to the plume and ionises part of the neutral shell driven by the nebula. This shell traps the ionising radiation, slowing the emergence of a rapid ``champagne'' flow of photoionised gas.}
	\label{fig:diagram}
\end{figure}

The goal of this paper is to understand how winds shape \HII regions in detail by simulating an object similar to the Orion nebula (M42) and exploring a set of physical parameters that influence such systems. The goal is to establish how robust our results are to changes in numerical and physical recipes, to identify whether there are significant gaps in our basic understanding of how \HII regions evolve, and thus to inform how more complex systems should be tackled. We self-consistently form a star similar to the ionising source of M42 ($\theta^1$ Ori C) inside a turbulent cloud. We vary the feedback processes included, physical recipes for the treatment of interfaces in the \HII region, and the random seed used to generate the initial conditions. 

The surprising outcome of this effect is that feedback becomes \textit{less} effective when winds are included. \rev{This is because the ionising photon budget is reduced by recombination, which occurs at a rate roughly proportional to density squared. The recombination rate in a dense, wind-swept shell, is thus higher than a typical density profile initially around a young star that decreases with radius. The dense shell is thus found to delay the rapid expansion of ionisation fronts around young stars, in agreement with the analytic solutions found in \cite{Geen2022}.}

We discuss whether we reproduce observational metrics for M42, including a dense shell containing neutral (non-photoionised) hydrogen \citep{Pabst2019}, an embedded wind bubble \citep{Guedel2008} and a similar radial expansion (a maximum extent of 4 pc with an age of roughly 200 kyr; see \citealp{Pabst2020}). We are to first order broadly successful with these criteria. A key aspect missing from previous work is the presence of a neutral shell rather than a rapidly-escaping champagne flow. We argue that champagne flows can still occur, but mainly when the \HII region encounters a sharp density discontinuity as in \cite{TenorioTagle1979} and \cite{Whitworth1979}. Champagne flows stemming from steep density gradients \citep{Franco1990} can effectively be trapped by the shell around the wind bubble \citep[see][]{Geen2022}. We plot a schematic depicting the general structure of the resulting nebula in Figure \ref{fig:diagram}.

We find that even at our small scales, the wind bubble is highly turbulent, with strong eddy mixing leading to thermal pressure losses in the hot wind-shocked gas. To correct for numerical overcooling of expanding shock-fronts in static grids \citep{Fierlinger2016,Gentry2017}, we run simulations with a mask over the contact discontinuity between the hot wind bubble and the gas outside and turn off cooling inside the mask. We also add a criterion that increases the resolution on pressure gradients, i.e. in the interface around the \HII region. The cooling mask does reduce cooling, with the energy in the wind bubble increased by a factor of a few, but not to completely adiabatic levels, as argued from observational estimates by, e.g., \cite{Pabst2019}. Instead, we argue the expansion rate of a wind bubble away from the star can be increased if the wind bubble is not spherical, but rather composed of a narrow chimney with a wider plume of hot wind-shocked gas at the edge of the \HII region.

Finally, some of our simulations with different initial random seeds exhibit wind bubbles that shrink or disappear entirely. We track a simulation that does this, and find that a form of catastrophic mixing occurs once a champagne flow breaks out of the cloud, forming a ``hot champagne'' flow, where the wind bubble as evidenced by gas above $10^6~$K disappears entirely. In one of our simulations a wind bubble struggles to form, likely because the environment it forms in is dense enough for the simulation to fall below resolution requirements set by \cite{Pittard2021}. This raises a difficult prospect for resolving stellar winds with consistent accuracy in molecular cloud simulations. Despite this, the energetics of the wind bubbles do show some convergence in 3 of the 5 seeds studied.

Our work succeeds in reproducing some of the observed features of nebulae such as M42 in Orion, including an extended wind bubble surrounded by a dense shell not overrun by a champagne flow. However, the influence of winds in the \HII region is highly non-linear, restricting the expansion of the photoionised region and displaying turbulent mixing. This adds to the overall cost of simulations that wish to reproduce stellar feedback in a molecular cloud environment. Further work remains to explore a wider parameter space relevant to star formation in galaxies.

\section{Data Availability}
\label{data_availability}
 
The data products associated with this paper, including simulation code, initial conditions, analysis code, intermediate data products and, given adequate storage, the full simulation outputs, can be made available by the authors upon reasonable request.

\section*{Acknowledgements}

The authors would like to thank Xander Tielens, Lex Kaper, Ben Keller and Eric Pellegrini for useful discussions during the writing of this paper. \rev{The authors would additionally like to thank the anonymous referee for their careful and detailed comments that helped improve the clarity of the manuscript}.
The simulations in this paper were performed on the Dutch National Supercomputing cluster Snellius at SURFsara. The authors gratefully acknowledge the SURFsara ResearchDrive facility for remote data sharing. We gratefully acknowledge the Common Computing Facility (CCF) of the LABEX Lyon Institute of Origins (ANR-10 LABX-0066). Copies of the data were also stored on the data storage service SDS@hd supported by the Ministry of Science, Research and the Arts Baden-W\"urttemberg (MWK) and the German Research Foundation (DFG) through grant INST 35/1314-1 FUGG. SG acknowledges support from a NOVA grant for the theory of massive star formation and a Spinoza award of the Dutch Science Organization (NWO) for research on the physics and chemistry of the interstellar medium. TK was supported by the National Research Foundation of Korea (NRF-2020R1C1C1007079 and 2022R1A6A1A03053472).



\bibliographystyle{mnras}
\bibliography{samgeen} 

\begin{thebibliography}{}
\makeatletter
\relax
\def\mn@urlcharsother{\let\do\@makeother \do\$\do\&\do\#\do\^\do\_\do\%\do\~}
\def\mn@doi{\begingroup\mn@urlcharsother \@ifnextchar [ {\mn@doi@}
  {\mn@doi@[]}}
\def\mn@doi@[#1]#2{\def\@tempa{#1}\ifx\@tempa\@empty \href
  {http://dx.doi.org/#2} {doi:#2}\else \href {http://dx.doi.org/#2} {#1}\fi
  \endgroup}
\def\mn@eprint#1#2{\mn@eprint@#1:#2::\@nil}
\def\mn@eprint@arXiv#1{\href {http://arxiv.org/abs/#1} {{\tt arXiv:#1}}}
\def\mn@eprint@dblp#1{\href {http://dblp.uni-trier.de/rec/bibtex/#1.xml}
  {dblp:#1}}
\def\mn@eprint@#1:#2:#3:#4\@nil{\def\@tempa {#1}\def\@tempb {#2}\def\@tempc
  {#3}\ifx \@tempc \@empty \let \@tempc \@tempb \let \@tempb \@tempa \fi \ifx
  \@tempb \@empty \def\@tempb {arXiv}\fi \@ifundefined
  {mn@eprint@\@tempb}{\@tempb:\@tempc}{\expandafter \expandafter \csname
  mn@eprint@\@tempb\endcsname \expandafter{\@tempc}}}

\bibitem[\protect\citeauthoryear{Ali, Harries  \& Douglas}{Ali
  et~al.}{2018}]{Ali2018}
Ali A.,  Harries T.~J.,   Douglas T.~A.,  2018, \mn@doi [Monthly Notices of the
  Royal Astronomical Society] {10.1093/mnras/sty1001}, 477, 5422

\bibitem[\protect\citeauthoryear{Ali, Bending  \& Dobbs}{Ali
  et~al.}{2022}]{Ali2022}
Ali A.~A.,  Bending T. J.~R.,   Dobbs C.~L.,  2022, \mn@doi [Monthly Notices of
  the Royal Astronomical Society] {10.1093/mnras/stac025}, 510, 5592

\bibitem[\protect\citeauthoryear{Alves \& Bouy}{Alves \&
  Bouy}{2012}]{Alves2012}
Alves J.,  Bouy H.,  2012, \mn@doi [Astronomy and Astrophysics]
  {10.1051/0004-6361/201220119}, 547

\bibitem[\protect\citeauthoryear{Audit \& Hennebelle}{Audit \&
  Hennebelle}{2005}]{Audit2005}
Audit E.,  Hennebelle P.,  2005, \mn@doi [Astronomy and Astrophysics]
  {10.1051/0004-6361:20041474}, 433, 1

\bibitem[\protect\citeauthoryear{Balega, Chentsov, Leushin, Rzaev  \&
  Weigelt}{Balega et~al.}{2014}]{Balega2014}
Balega Y.~Y.,  Chentsov E.~L.,  Leushin V.~V.,  Rzaev A.~K.,   Weigelt G.,
  2014, \mn@doi [Astrophysical Bulletin] {10.1134/S1990341314010052}, 69, 46

\bibitem[\protect\citeauthoryear{Barnes, Longmore, Dale, Krumholz, Kruijssen
  \& Bigiel}{Barnes et~al.}{2020}]{Barnes2020}
Barnes A.~T.,  Longmore S.~N.,  Dale J.~E.,  Krumholz M.~R.,  Kruijssen
  J.~M.~D.,   Bigiel F.,  2020, \mn@doi [Monthly Notices of the Royal
  Astronomical Society] {10.1093/mnras/staa2719}, 498, 4906

\bibitem[\protect\citeauthoryear{Bate}{Bate}{2019}]{Bate2019}
Bate M.~R.,  2019, \mn@doi [Monthly Notices of the Royal Astronomical Society]
  {10.1093/mnras/stz103}, 484, 2341

\bibitem[\protect\citeauthoryear{Beuther et~al.,}{Beuther
  et~al.}{2022}]{Beuther2022}
Beuther H.,  et~al., 2022, Astronomy \& Astrophysics, 659, A77

\bibitem[\protect\citeauthoryear{Bieri, Naab, Geen, Coles, Pakmor  \&
  Walch}{Bieri et~al.}{2023}]{Bieri2023}
Bieri R.,  Naab T.,  Geen S.,  Coles J.~P.,  Pakmor R.,   Walch S.,  2023,
  Monthly Notices of the Royal Astronomical Society, 523, 6336

\bibitem[\protect\citeauthoryear{Bleuler \& Teyssier}{Bleuler \&
  Teyssier}{2014}]{Bleuler2014a}
Bleuler A.,  Teyssier R.,  2014, \mn@doi [Monthly Notices of the Royal
  Astronomical Society] {10.1093/mnras/stu2005}, 445, 4015

\bibitem[\protect\citeauthoryear{Bleuler, Teyssier, Carassou  \&
  Martizzi}{Bleuler et~al.}{2014}]{Bleuler2014}
Bleuler A.,  Teyssier R.,  Carassou S.,   Martizzi D.,  2014, \mn@doi
  [Computational Astrophysics and Cosmology] {10.1186/s40668-015-0009-7}, 2, 16

\bibitem[\protect\citeauthoryear{Brown, de Geus  \& de Zeeuw}{Brown
  et~al.}{1994}]{Brown1994}
Brown A. G.~A.,  de Geus E.~J.,   de Zeeuw P.~T.,  1994, Astronomy \&
  Astrophysics, 289, 101

\bibitem[\protect\citeauthoryear{Chevance, Krumholz, McLeod, Ostriker,
  Rosolowsky  \& Sternberg}{Chevance et~al.}{2022}]{Chevance2022a}
Chevance M.,  Krumholz M.~R.,  McLeod A.~F.,  Ostriker E.~C.,  Rosolowsky
  E.~W.,   Sternberg A.,  2022, arXiv e-prints, p. arXiv:2203.09570

\bibitem[\protect\citeauthoryear{Comeron}{Comeron}{1997}]{Comeron1997}
Comeron F.,  1997, Astronomy \& Astrophysics, 326, 1195

\bibitem[\protect\citeauthoryear{Dale, Bonnell, Clarke  \& Bate}{Dale
  et~al.}{2005}]{Dale2005}
Dale J.~E.,  Bonnell I.~A.,  Clarke C.~J.,   Bate M.~R.,  2005, \mn@doi
  [Monthly Notices of the Royal Astronomical Society]
  {10.1111/j.1365-2966.2005.08806.x}, 358, 291

\bibitem[\protect\citeauthoryear{Dale, Ercolano  \& Bonnell}{Dale
  et~al.}{2012}]{Dale2012}
Dale J.~E.,  Ercolano B.,   Bonnell I.~A.,  2012, \mn@doi [Monthly Notices of
  the Royal Astronomical Society] {10.1111/j.1365-2966.2012.21205.x}, 424, 377

\bibitem[\protect\citeauthoryear{Dale, Ngoumou, Ercolano  \& Bonnell}{Dale
  et~al.}{2014}]{Dale2014}
Dale J.~E.,  Ngoumou J.,  Ercolano B.,   Bonnell I.~A.,  2014, \mn@doi [Monthly
  Notices of the Royal Astronomical Society] {10.1093/mnras/stu816}, 442, 694

\bibitem[\protect\citeauthoryear{Draine}{Draine}{2011}]{Draine2011}
Draine B.~T.,  2011, \mn@doi [The Astrophysical Journal]
  {10.1088/0004-637X/732/2/100}, 732, 100

\bibitem[\protect\citeauthoryear{Drass, Haas, Chini, Bayo, Hackstein,
  Hoffmeister, Godoy  \& Vogt}{Drass et~al.}{2016}]{Drass2016}
Drass H.,  Haas M.,  Chini R.,  Bayo A.,  Hackstein M.,  Hoffmeister V.,  Godoy
  N.,   Vogt N.,  2016, \mn@doi [Monthly Notices of the Royal Astronomical
  Society] {10.1093/mnras/stw1094}, 461, 1734

\bibitem[\protect\citeauthoryear{Dwarkadas}{Dwarkadas}{2022}]{Dwarkadas2022}
Dwarkadas V.~V.,  2022, arXiv e-prints, p. arXiv:2202.09432

\bibitem[\protect\citeauthoryear{Ekström et~al.,}{Ekström
  et~al.}{2012}]{Ekstrom2012}
Ekström S.,  et~al., 2012, \mn@doi [Astronomy \& Astrophysics]
  {10.1051/0004-6361/201117751}, 537, A146

\bibitem[\protect\citeauthoryear{Ferland}{Ferland}{2003}]{Ferland2003}
Ferland G.~J.,  2003, \mn@doi [Annual Review of Astronomy and Astrophysics]
  {10.1146/annurev.astro.41.011802.094836}, 41, 517

\bibitem[\protect\citeauthoryear{Fielding, Ostriker, Bryan  \& Jermyn}{Fielding
  et~al.}{2020}]{Fielding2020}
Fielding D.~B.,  Ostriker E.~C.,  Bryan G.~L.,   Jermyn A.~S.,  2020, \mn@doi
  [Astrophysical Journal Letters] {10.3847/2041-8213/ab8d2c}, 894, L24

\bibitem[\protect\citeauthoryear{Fierlinger, Burkert, Ntormousi, Fierlinger,
  Schartmann, Ballone, Krause  \& Diehl}{Fierlinger
  et~al.}{2016}]{Fierlinger2016}
Fierlinger K.~M.,  Burkert A.,  Ntormousi E.,  Fierlinger P.,  Schartmann M.,
  Ballone A.,  Krause M. G.~H.,   Diehl R.,  2016, \mn@doi [Monthly Notices of
  the Royal Astronomical Society] {10.1093/mnras/stv2699}, 456, 710

\bibitem[\protect\citeauthoryear{Franco, Tenorio-Tagle  \& Bodenheimer}{Franco
  et~al.}{1990}]{Franco1990}
Franco J.,  Tenorio-Tagle G.,   Bodenheimer P.,  1990, \mn@doi [The
  Astrophysical Journal] {10.1086/168300}, 349, 126

\bibitem[\protect\citeauthoryear{Fromang, Hennebelle  \& Teyssier}{Fromang
  et~al.}{2006}]{Fromang2006}
Fromang S.,  Hennebelle P.,   Teyssier R.,  2006, \mn@doi [Astronomy \&
  Astrophysics] {10.1051/0004-6361:20065371}, 457, 371

\bibitem[\protect\citeauthoryear{Gatto et~al.,}{Gatto et~al.}{2017}]{Gatto2017}
Gatto A.,  et~al., 2017, \mn@doi [Monthly Notices of the Royal Astronomical
  Society] {10.1093/mnras/stw3209}, 466, 1903

\bibitem[\protect\citeauthoryear{Geen \& de Koter}{Geen \&
  de~Koter}{2022}]{Geen2022}
Geen S.,  de Koter A.,  2022, Monthly Notices of the Royal Astronomical
  Society, 509, pp.4498

\bibitem[\protect\citeauthoryear{Geen, Soler  \& Hennebelle}{Geen
  et~al.}{2017}]{Geen2017}
Geen S.,  Soler J.~D.,   Hennebelle P.,  2017, Monthly Notices of the Royal
  Astronomical Society, 471, 4844

\bibitem[\protect\citeauthoryear{Geen, Watson, Rosdahl, Bieri, Klessen  \&
  Hennebelle}{Geen et~al.}{2018}]{Geen2018}
Geen S.,  Watson S.~K.,  Rosdahl J.,  Bieri R.,  Klessen R.~S.,   Hennebelle
  P.,  2018, Monthly Notices of the Royal Astronomical Society, 481, 2548

\bibitem[\protect\citeauthoryear{Geen, Bieri, Rosdahl  \& de Koter}{Geen
  et~al.}{2021}]{Geen2020}
Geen S.,  Bieri R.,  Rosdahl J.,   de Koter A.,  2021, Monthly Notices of the
  Royal Astronomical Society, 501, 1352

\bibitem[\protect\citeauthoryear{Gentry, Krumholz, Dekel  \& Madau}{Gentry
  et~al.}{2017}]{Gentry2017}
Gentry E.~S.,  Krumholz M.~R.,  Dekel A.,   Madau P.,  2017, Monthly Notices of
  the Royal Astronomical Society, 465, 2471

\bibitem[\protect\citeauthoryear{Georgy, Ekström, Meynet, Massey, Levesque,
  Hirschi, Eggenberger  \& Maeder}{Georgy et~al.}{2012}]{Georgy2012}
Georgy C.,  Ekström S.,  Meynet G.,  Massey P.,  Levesque E.~M.,  Hirschi R.,
  Eggenberger P.,   Maeder A.,  2012, \mn@doi [Astronomy \& Astrophysics]
  {10.1051/0004-6361/201118340}, 542

\bibitem[\protect\citeauthoryear{Girichidis et~al.,}{Girichidis
  et~al.}{2020}]{Girichidis2020}
Girichidis P.,  et~al., 2020, \mn@doi [Space Science Reviews]
  {10.1007/s11214-020-00693-8}, 216, 68

\bibitem[\protect\citeauthoryear{Gritschneder, Naab, Walch, Burkert  \&
  Heitsch}{Gritschneder et~al.}{2009}]{Gritschneder2009}
Gritschneder M.,  Naab T.,  Walch S.,  Burkert A.,   Heitsch F.,  2009, \mn@doi
  [The Astrophysical Journal] {10.1088/0004-637X/694/1/L26}, 694, L26

\bibitem[\protect\citeauthoryear{Grudić, Guszejnov, Offner, Rosen, Raju,
  Faucher-Giguère  \& Hopkins}{Grudić et~al.}{2022}]{Grudic2022}
Grudić M.~Y.,  Guszejnov D.,  Offner S. S.~R.,  Rosen A.~L.,  Raju A.~N.,
  Faucher-Giguère C.-A.,   Hopkins P.~F.,  2022, \mn@doi [Monthly Notices of
  the Royal Astronomical Society] {10.1093/mnras/stac526}

\bibitem[\protect\citeauthoryear{Guedel, Briggs, Montmerle, Audard, Rebull  \&
  Skinner}{Guedel et~al.}{2008}]{Guedel2008}
Guedel M.,  Briggs K.~R.,  Montmerle T.,  Audard M.,  Rebull L.,   Skinner
  S.~L.,  2008, \mn@doi [Science] {10.1126/science.1149926}, 319, 309

\bibitem[\protect\citeauthoryear{Guszejnov, Grudić, Offner, Faucher-Giguère,
  Hopkins  \& Rosen}{Guszejnov et~al.}{2022}]{Guszejnov2022}
Guszejnov D.,  Grudić M.~Y.,  Offner S.~S.,  Faucher-Giguère C.~A.,  Hopkins
  P.~F.,   Rosen A.~L.,  2022, \mn@doi [Monthly Notices of the Royal
  Astronomical Society] {10.1093/MNRAS/STAC2060}, 515, 4929

\bibitem[\protect\citeauthoryear{Kahn}{Kahn}{1954}]{Kahn1954}
Kahn F.~D.,  1954, Bulletin of the Astronomical Institutes of the Netherlands,
  12, 187

\bibitem[\protect\citeauthoryear{Kahn}{Kahn}{1980}]{Kahn1980}
Kahn F.~D.,  1980, Astronomy and Astrophysics, 83, 303

\bibitem[\protect\citeauthoryear{Kavak, Goicoechea, Pabst, Bally, van~der Tak
  \& Tielens}{Kavak et~al.}{2022}]{Kavak2022}
Kavak U.,  Goicoechea J.~R.,  Pabst C.~H.~M.,  Bally J.,  van~der Tak F.~F.~S.,
    Tielens A.~G.~G.~M.,  2022, arXiv e-prints, p. arXiv:2202.04711

\bibitem[\protect\citeauthoryear{Kimm, Bieri, Geen, Rosdahl, Blaizot,
  Michel-Dansac  \& Garel}{Kimm et~al.}{2022}]{kimm2022}
Kimm T.,  Bieri R.,  Geen S.,  Rosdahl J.,  Blaizot J.,  Michel-Dansac L.,
  Garel T.,  2022, \mn@doi [The Astrophysical Journal Supplement Series]
  {10.3847/1538-4365/ac426d}, 259, 21

\bibitem[\protect\citeauthoryear{Klessen, Heitsch  \& Mac~Low}{Klessen
  et~al.}{2000}]{Klessen2000}
Klessen R.~S.,  Heitsch F.,   Mac~Low M.-M.,  2000, \mn@doi [The Astrophysical
  Journal] {10.1086/308891}, 535, 887

\bibitem[\protect\citeauthoryear{Kraus et~al.,}{Kraus et~al.}{2009}]{Kraus2009}
Kraus S.,  et~al., 2009, \mn@doi [Astronomy and Astrophysics]
  {10.1051/0004-6361/200810368}, 497, 195

\bibitem[\protect\citeauthoryear{Kruijssen et~al.,}{Kruijssen
  et~al.}{2019}]{Kruijssen2019}
Kruijssen J. M.~D.,  et~al., 2019, \mn@doi [Nature]
  {10.1038/s41586-019-1194-3}, 569, 519

\bibitem[\protect\citeauthoryear{Kuiper \& Hosokawa}{Kuiper \&
  Hosokawa}{2018}]{Kuiper2018}
Kuiper R.,  Hosokawa T.,  2018, Astronomy \& Astrophysics, 616, 22

\bibitem[\protect\citeauthoryear{Lancaster, Ostriker, Kim  \& Kim}{Lancaster
  et~al.}{2021a}]{Lancaster2021a}
Lancaster L.,  Ostriker E.~C.,  Kim J.-G.,   Kim C.-G.,  2021a, \mn@doi [The
  Astrophysical Journal] {10.3847/1538-4357/abf8ab}, 914, 89

\bibitem[\protect\citeauthoryear{Lancaster, Ostriker, Kim  \& Kim}{Lancaster
  et~al.}{2021b}]{Lancaster2021b}
Lancaster L.,  Ostriker E.~C.,  Kim J.-G.,   Kim C.-G.,  2021b, \mn@doi [The
  Astrophysical Journal] {10.3847/1538-4357/abf8ac}, 914, 90

\bibitem[\protect\citeauthoryear{Lancaster, Ostriker, Kim  \& Kim}{Lancaster
  et~al.}{2021c}]{Lancaster2021c}
Lancaster L.,  Ostriker E.~C.,  Kim J.-G.,   Kim C.-G.,  2021c, \mn@doi
  [Astrophysical Journal Letters] {10.3847/2041-8213/ac3333}, 922, L3

\bibitem[\protect\citeauthoryear{Lebreuilly, Commerçon  \& Laibe}{Lebreuilly
  et~al.}{2019}]{Lebreuilly2019}
Lebreuilly U.,  Commerçon B.,   Laibe G.,  2019, \mn@doi [Astronomy \&
  Astrophysics] {10.1051/0004-6361/201834147}, 626, A96

\bibitem[\protect\citeauthoryear{Leitherer, Ekström, Meynet, Schaerer, Agienko
   \& Levesque}{Leitherer et~al.}{2014}]{Leitherer2014}
Leitherer C.,  Ekström S.,  Meynet G.,  Schaerer D.,  Agienko K.~B.,
  Levesque E.~M.,  2014, \mn@doi [The Astrophysical Journal Supplement Series]
  {10.1088/0067-0049/212/1/14}, 212, 14

\bibitem[\protect\citeauthoryear{Mac~Low \& McCray}{Mac~Low \&
  McCray}{1988}]{MacLow1988}
Mac~Low M.-M.,  McCray R.,  1988, \mn@doi [The Astrophysical Journal]
  {10.1086/165936}, 324, 776

\bibitem[\protect\citeauthoryear{Mathews}{Mathews}{1967}]{Mathews1967}
Mathews W.~G.,  1967, \mn@doi [The Astrophysical Journal] {10.1086/149087},
  147, 965

\bibitem[\protect\citeauthoryear{McKee \& Ostriker}{McKee \&
  Ostriker}{1977}]{McKee1977}
McKee C.~F.,  Ostriker J.~P.,  1977, \mn@doi [The Astrophysical Journal]
  {10.1086/155667}, 218, 148

\bibitem[\protect\citeauthoryear{Miyoshi \& Kusano}{Miyoshi \&
  Kusano}{2005}]{Miyoshi2005}
Miyoshi T.,  Kusano K.,  2005, \mn@doi [Journal of Computational Physics]
  {10.1016/j.jcp.2005.02.017}, 208, 315

\bibitem[\protect\citeauthoryear{Morlino, Blasi, Peretti  \&
  Cristofari}{Morlino et~al.}{2021}]{Morlino2021}
Morlino G.,  Blasi P.,  Peretti E.,   Cristofari P.,  2021, \mn@doi [Monthly
  Notices of the Royal Astronomical Society] {10.1093/mnras/stab690}, 504, 6096

\bibitem[\protect\citeauthoryear{O'Dell, Kollatschny  \& Ferland}{O'Dell
  et~al.}{2017}]{ODell2017}
O'Dell C.~R.,  Kollatschny W.,   Ferland G.~J.,  2017, \mn@doi [The
  Astrophysical Journal] {10.3847/1538-4357/aa6198}, 837, 151

\bibitem[\protect\citeauthoryear{Olivier, Lopez, Rosen, Nayak, Reiter, Krumholz
   \& Bolatto}{Olivier et~al.}{2021}]{Oliver2021}
Olivier G.~M.,  Lopez L.~A.,  Rosen A.~L.,  Nayak O.,  Reiter M.,  Krumholz
  M.~R.,   Bolatto A.~D.,  2021, \mn@doi [The Astrophysical Journal]
  {10.3847/1538-4357/abd24a}, 908, 68

\bibitem[\protect\citeauthoryear{Pabst et~al.,}{Pabst et~al.}{2019}]{Pabst2019}
Pabst C.,  et~al., 2019, \mn@doi [Nature] {10.1038/s41586-018-0844-1}, 565, 618

\bibitem[\protect\citeauthoryear{Pabst et~al.,}{Pabst et~al.}{2020}]{Pabst2020}
Pabst C. H.~M.,  et~al., 2020, \mn@doi [Astronomy \& Astrophysics]
  {10.1051/0004-6361/202037560}, 639, A2

\bibitem[\protect\citeauthoryear{Peters, Banerjee, Klessen, Mac~Low,
  Galván-Madrid  \& Keto}{Peters et~al.}{2010}]{Peters2010}
Peters T.,  Banerjee R.,  Klessen R.~S.,  Mac~Low M.-M.,  Galván-Madrid R.,
  Keto E.~R.,  2010, \mn@doi [The Astrophysical Journal]
  {10.1088/0004-637X/711/2/1017}, 711, 1017

\bibitem[\protect\citeauthoryear{Pittard, Wareing  \& Kupilas}{Pittard
  et~al.}{2021}]{Pittard2021}
Pittard J.~M.,  Wareing C.~J.,   Kupilas M.~M.,  2021, \mn@doi [Monthly Notices
  of the Royal Astronomical Society] {10.1093/mnras/stab2712}, 508, 1768

\bibitem[\protect\citeauthoryear{Puls, Sundqvist  \& Markova}{Puls
  et~al.}{2015}]{Puls2015}
Puls J.,  Sundqvist J.~O.,   Markova N.,  2015. Cambridge University Press, pp
  25--36, \mn@doi{10.1017/S174392131400622X}

\bibitem[\protect\citeauthoryear{Rahner, Pellegrini, Glover  \& Klessen}{Rahner
  et~al.}{2017}]{Rahner2017}
Rahner D.,  Pellegrini E.~W.,  Glover S. C.~O.,   Klessen R.~S.,  2017, Monthly
  Notices of the Royal Astronomical Society, 470, 4453

\bibitem[\protect\citeauthoryear{Rathjen et~al.,}{Rathjen
  et~al.}{2021}]{Rathjen2021}
Rathjen T.-E.,  et~al., 2021, \mn@doi [Monthly Notices of the Royal
  Astronomical Society] {10.1093/mnras/stab900}, 504, 1039

\bibitem[\protect\citeauthoryear{Rey-Raposo, Dobbs, Agertz  \& Alig}{Rey-Raposo
  et~al.}{2017}]{Rey-Raposo2017}
Rey-Raposo R.,  Dobbs C.,  Agertz O.,   Alig C.,  2017, \mn@doi [Monthly
  Notices of the Royal Astronomical Society] {10.1093/mnras/stw2607}, 464, 3536

\bibitem[\protect\citeauthoryear{Rio, Robberto, Hillenbrand, Henning  \&
  Stassun}{Rio et~al.}{2012}]{DaRio2012}
Rio N.~D.,  Robberto M.,  Hillenbrand L.~A.,  Henning T.,   Stassun K.~G.,
  2012, \mn@doi [The Astrophysical Journal] {10.1088/0004-637X/748/1/14}, 748,
  14

\bibitem[\protect\citeauthoryear{Rogers \& Pittard}{Rogers \&
  Pittard}{2013}]{Rogers2013}
Rogers H.,  Pittard J.~M.,  2013, \mn@doi [Monthly Notices of the Royal
  Astronomical Society] {10.1093/mnras/stt255}, 431, 1337

\bibitem[\protect\citeauthoryear{Rosdahl \& Teyssier}{Rosdahl \&
  Teyssier}{2015}]{Rosdahl2015}
Rosdahl J.,  Teyssier R.,  2015, \mn@doi [Monthly Notices of the Royal
  Astronomical Society] {10.1093/mnras/stv567}, 449, 4380

\bibitem[\protect\citeauthoryear{Rosdahl, Blaizot, Aubert, Stranex  \&
  Teyssier}{Rosdahl et~al.}{2013}]{Rosdahl2013}
Rosdahl J.,  Blaizot J.,  Aubert D.,  Stranex T.,   Teyssier R.,  2013, \mn@doi
  [Monthly Notices of the Royal Astronomical Society] {10.1093/mnras/stt1722},
  436, 2188

\bibitem[\protect\citeauthoryear{Rosdahl et~al.,}{Rosdahl
  et~al.}{2018}]{Rosdahl2018}
Rosdahl J.,  et~al., 2018, Monthly Notices of the Royal Astronomical Society,
  479, 994

\bibitem[\protect\citeauthoryear{Rosen, Lopez, Krumholz  \& Ramirez-Ruiz}{Rosen
  et~al.}{2014}]{Rosen2014}
Rosen A.~L.,  Lopez L.~A.,  Krumholz M.~R.,   Ramirez-Ruiz E.,  2014, \mn@doi
  [Monthly Notices of the Royal Astronomical Society] {10.1093/mnras/stu1037},
  442, 2701

\bibitem[\protect\citeauthoryear{Rosen, Offner, Foley  \& Lopez}{Rosen
  et~al.}{2021}]{Rosen2021}
Rosen A.~L.,  Offner S. S.~R.,  Foley M.~M.,   Lopez L.~A.,  2021, arXiv
  e-prints, p. arXiv:2107.12397

\bibitem[\protect\citeauthoryear{Schneider et~al.,}{Schneider
  et~al.}{2020}]{Schneider2020}
Schneider N.,  et~al., 2020, \mn@doi [Publications of the Astronomical Society
  of the Pacific] {10.1088/1538-3873/aba840}, 132, 104301

\bibitem[\protect\citeauthoryear{Simón-Díaz, Herrero, Esteban  \&
  Najarro}{Simón-Díaz et~al.}{2006}]{SimonDiaz2006}
Simón-Díaz S.,  Herrero A.,  Esteban C.,   Najarro F.,  2006, \mn@doi
  [Astronomy \& Astrophysics] {10.1051/0004-6361:20053066}, 448, 351

\bibitem[\protect\citeauthoryear{Spitzer}{Spitzer}{1962}]{Spitzer1962}
Spitzer L.,  1962, Physics of Fully Ionized Gases, 2nd edn. edn.
New York, Interscience Publishers

\bibitem[\protect\citeauthoryear{Sutherland \& Dopita}{Sutherland \&
  Dopita}{1993}]{Sutherland1993}
Sutherland R.~S.,  Dopita M.~A.,  1993, \mn@doi [The Astrophysical Journal
  Supplement Series] {10.1086/191823}, 88, 253

\bibitem[\protect\citeauthoryear{Tan, Oh  \& Gronke}{Tan
  et~al.}{2021}]{Tan2021}
Tan B.,  Oh S.~P.,   Gronke M.,  2021, \mn@doi [Monthly Notices of the Royal
  Astronomical Society] {10.1093/mnras/stab053}, 502, 3179

\bibitem[\protect\citeauthoryear{Tenorio-Tagle}{Tenorio-Tagle}{1979}]{TenorioTagle1979}
Tenorio-Tagle G.,  1979, Astronomy and Astrophysics, 71, 59

\bibitem[\protect\citeauthoryear{Teyssier}{Teyssier}{2002}]{Teyssier2002}
Teyssier R.,  2002, \mn@doi [Astronomy \& Astrophysics]
  {10.1051/0004-6361:20011817}, 385, 337

\bibitem[\protect\citeauthoryear{Tumlinson, Peeples  \& Werk}{Tumlinson
  et~al.}{2017}]{Tumlinson2017}
Tumlinson J.,  Peeples M.~S.,   Werk J.~K.,  2017, \mn@doi [Annual Review of
  Astronomy and Astrophysics] {10.1146/annurev-astro-091916-055240}, 55, 389

\bibitem[\protect\citeauthoryear{Verliat, Hennebelle, González, Lee  \&
  Geen}{Verliat et~al.}{2022}]{Verliat2022}
Verliat A.,  Hennebelle P.,  González M.,  Lee Y.-N.,   Geen S.,  2022, arXiv
  e-prints, p. arXiv:2202.02237

\bibitem[\protect\citeauthoryear{Vink, Muijres, Anthonisse, de Koter, Graefener
   \& Langer}{Vink et~al.}{2011}]{Vink2011}
Vink J.~S.,  Muijres L.~E.,  Anthonisse B.,  de Koter A.,  Graefener G.,
  Langer N.,  2011, \mn@doi [Astronomy \& Astrophysics]
  {10.1051/0004-6361/201116614}, 531, 132

\bibitem[\protect\citeauthoryear{Walch, Whitworth, Bisbas, Wünsch  \&
  Hubber}{Walch et~al.}{2012}]{Walch2012}
Walch S.~K.,  Whitworth A.~P.,  Bisbas T.,  Wünsch R.,   Hubber D.,  2012,
  \mn@doi [Monthly Notices of the Royal Astronomical Society]
  {10.1111/j.1365-2966.2012.21767.x}, 427, 625

\bibitem[\protect\citeauthoryear{Weaver, McCray, Castor, Shapiro  \&
  Moore}{Weaver et~al.}{1977}]{Weaver1977}
Weaver R.,  McCray R.,  Castor J.,  Shapiro P.,   Moore R.,  1977, \mn@doi [The
  Astrophysical Journal] {10.1086/155692}, 218, 377

\bibitem[\protect\citeauthoryear{Whitworth}{Whitworth}{1979}]{Whitworth1979}
Whitworth A.,  1979, Monthly Notices of the Royal Astronomical Society, 186, 59

\makeatother
\end{thebibliography}




%
%
%


\bsp	
\label{lastpage}
\end{document}